\documentclass[traditabstract]{aa} %

\usepackage{graphicx}
\usepackage{epstopdf}
\usepackage{color}
\usepackage{savesym}
\savesymbol{rotate}
\usepackage{deluxetable} 
\restoresymbol{DTBL}{rotate}
\usepackage{times,epsfig} 
\usepackage{mathrsfs}  
\usepackage{natbib}
\usepackage{amssymb}
\bibpunct{(}{)}{;}{a}{}{,} 
\usepackage[english]{babel}
\usepackage{longtable}
\usepackage[english]{babel}
\usepackage{url}
\usepackage{hyperref}
\usepackage[normalem]{ulem}

\def\ifundefined#1{\expandafter\ifx\csname#1\endcsname\relax}
\ifundefined{ensuremath}\def\ensuremath#1{\relax\ifmmode{#1}}
\else${#1}$\fi\else\relax\fi
\ifundefined{nuc}\def\nuc#1#2{\relax\ifmmode{}^{#1}{\protect\mathrm{#2}}
\else${}^{#1}$#2\fi}\else\relax\fi

\usepackage{morefloats}
\usepackage{inputenc}

\begin{document}

\title{\textit{The Carnegie Supernova Project II.} Early observations and progenitor constraints of the Type~Ib supernova LSQ13abf\thanks{This paper includes data gathered  with the Nordic Optical Telescope  at the Observatorio del Roque de los Muchachos, La Palma, Spain, and the 6.5 meter Magellan Telescopes located at Las Campanas Observatory, Chile.}}

\author{M. D. Stritzinger\inst{1}
\and F. Taddia\inst{1}
\and S. Holmbo\inst{1}
\and E. Baron\inst{2,1} 
\and C. Contreras\inst{1,3}
\and E. Karamehmetoglu\inst{1,4}
\and M. M. Phillips\inst{3} 
\and J. Sollerman\inst{4}
\and N. B. Suntzeff\inst{5}
\and J. Vinko\inst{6,7,8}
\and C. Ashall\inst{9}
\and C. Avila\inst{3}
\and C. R. Burns\inst{10}
\and A. Campillay\inst{3}
\and S. Castellon\inst{3}
\and G. Folatelli\inst{11,12,13}
\and L. Galbany\inst{14}
\and P. Hoeflich\inst{9}
\and E. Y. Hsiao\inst{9}
\and G. H. Marion\inst{8}
\and N. Morrell\inst{3}
\and J. C.Wheeler\inst{8}
}

\institute{
Department of Physics and Astronomy, Aarhus University, 
Ny Munkegade 120, DK-8000 Aarhus C, Denmark  
\and
Homer L. Dodge Department of Physics and Astronomy, University of Oklahoma, 440 W. Brooks, Rm 100, Norman, OK 73019-2061, USA
\and
Carnegie Observatories, Las Campanas Observatory, Casilla 601, La Serena, Chile
\and
The Oskar Klein Centre, Department of Astronomy, Stockholm University, AlbaNova, 10691 Stockholm, Sweden
\and
George P. and Cynthia Woods Mitchell Institute for Fundamental Physics and Astronomy, Texas A\&M University, Department of Physics and Astronomy, College Station, TX, 77843, USA
\and
Konkoly Observatory, CSFK,
Konkoly-Thege M. ut 15-17, Budapest, Hungary
\and
Department of Optics and Quantum Electronics, University of Szeged, Dom ter 9, Szeged, Hungary
\and
Department of Astronomy, University of Texas, 1 University Station C1400, Austin, TX 78712, USA
\and
 Department of Physics, Florida State University, Tallahassee, FL 32306, USA
 \and
 Observatories of the Carnegie Institution for Science, 813 Santa Barbara St., Pasadena, CA 91101, USA
 \and 
 Instituto de Astrofısica de La Plata (IALP), CONICET, Paseo del bosque S/N, 1900, Argentina
\and
Facultad de Ciencias Astron\'{o}micas y Geof\'{i}sicas (FCAG), Universidad Nacional de La Plata (UNLP), Paseo del bosque S/N, 1900, Argentina
\and 
Kavli Institute for the Physics and Mathematics of the Universe, Todai Institutes for Advanced Study, University of Tokyo, 5-1-5 Kashiwanoha, Kashiwa, Chiba 277-8583, Japan
 \and
 Departamento de F\'isica Te\'orica y del Cosmos, Universidad de Granada, E-18071 Granada, Spain
}

\date{Received; accepted}

\abstract{Supernova LSQ13abf was discovered soon after explosion by the La Silla-QUEST Survey and  followed by the \textit{Carnegie Supernova Project II} at optical and near-IR (NIR) wavelengths.
Our analysis indicates LSQ13abf was discovered within two days of explosion and its first $\approx$10 days of evolution reveal  a $B$-band light curve  with an abrupt drop in luminosity.
Contemporaneously,  the $V$-band  light curve exhibits a rise towards a first peak and the $r$- and $i$-band light curves show no early peak.
The early light-curve evolution of LSQ13abf is reminiscent of the post explosion cooling phase observed in the Type~Ib SN~2008D, and the similarity between the two objects extends over weeks.
Spectroscopically, LSQ13abf  also resembles SN~2008D with  P~Cygni \ion{He}{i}  features that strengthen over time. 
Spectral energy distributions are constructed from  broad-band photometry, and by fitting black-body (BB) functions  a UVOIR light curve is constructed, and the underlying BB-temperature and BB-radius profiles are estimated. Explosion parameters are estimated by simultaneously fitting an Arnett model to the UVOIR light curve and the velocity evolution derived from spectral features, and a post-shock breakout cooling model to the first two epochs of the bolometric evolution. This combined model  suggests an explosion energy of $1.27\pm0.23\times10^{51}$ ergs,  a relatively high ejecta mass of 5.94$\pm$1.10~$M_{\odot}$, a $^{56}$Ni mass of 0.16$\pm$0.02~$M_{\odot}$, and a progenitor-star radius of 28.0$\pm$7.5~$R_{\odot}$.
The ejecta mass suggests the origins of LSQ13abf lie with a  $> 25~M_{\odot}$ zero-age-main-sequence mass progenitor and its  estimated radius is three and nine times larger than values estimated from the same analysis applied to observations of SN~2008D and SN~1999ex, respectively. 
Alternatively, comparison of hydrodynamical simulations  of $\gtrsim$ 20-25 $M_{\odot}$ zero-age-main-sequence progenitors that evolve  to  pre-supernova envelope masses of $\lesssim$10 $M_{\odot}$   and extended ($\sim$ 100 $R_{\odot}$) envelopes  also  broadly match the observations of LSQ13abf.} 

\authorrunning{Stritzinger, Taddia, et al.}
\titlerunning{Early observations of the Type~Ib supernova LSQ13abf}

\maketitle

\section{Introduction}
\label{sec:intro}
Core-collapse (CC) supernovae (SNe) mark the  demise of massive  stars. The characteristics of their progenitor stars can be measured directly from deep, high-spatial resolution pre-explosion images \citep{smartt15}, but more frequently indirectly by modeling their SN emission. The observations of SNe at different temporal phases provides different insights on the nature of their progenitors. For example, late-phase observations enable us to peer into the deepest layers of the SN ejecta  \citep[e.g.][]{jerkstrand17}, providing clues on potential asymmetries, and through spectral synthesis, an estimate of  the zero-age-main-sequence (ZAMS) mass of the progenitor \citep[e.g.,][]{mazzali17}.  
Early-time observations, on the other hand,  provide  information on the SN explosion parameters, 
This  includes the kinetic energy ($E_K$) of the explosion, the ejecta mass ($M_{ej}$), the  amount of $^{56}$Ni synthesized during the explosion, and the level of  mixing of the radioactive isotopes. Furthermore, if the observations begin in the hours to days after explosion and catch the cooling tail that follows shock breakout, one can estimate the  radius of the progenitor at the moment of its death \citep[e.g.,][]{rabinak11}.

With the advent of fast, non-targeted optical surveys, very early discoveries occur more frequently than in the past, revealing in some cases peculiar emission features related to the progenitor structure. Before the era of large surveys, SNe were typically observed in the optical while they rise to the main peak followed by a decline; however, it is becoming more common to observe an early peak at blue wavelengths prior to the rise to the main peak several weeks after the explosion. 
This early blue flux observed in CC SNe arises from the rapid cooling of the ejecta after shock breakout 
\citep[e.g.][]{chevalier92,chevalier08}. Capturing these ephemeral epochs offers an avenue to elucidate details on  a number of physical processes. In addition to shock break out, other factors may also be at play that shape the early light curve. These include  (i) the presence of any extended structure of the progenitor star \citep{bersten13,nakar14,piro15},  (ii) an excess of emission produced by shock interaction between the rapidly expanding SN ejecta with a companion \citep{kasen10}, (iii)  the distribution of  radioactive material  \citep{folatelli06,bersten13,noebauer17,taddia1811mnb}, and (iv)  emission from a magnetar \citep{kasen15}.

The most famous and well-studied SN~1987A was a  hydrogen-rich, peculiar CC Type~II SN  that presented the first evidence of an early optical peak  \citep{hamuy88}. This discovery was followed with the hydrogen-poor, stripped-envelope Type~IIb SN~1993J,  showing a prominent cooling tail post explosion \citep{richmond94,clocchiatti95}, and later by other Type~IIb SNe, e.g., SN~2011dh \citep{arcavi11}, SN~2011fu \citep{morales15}, and SN~2013df \citep{morales14}. 
 Even super-luminous SNe (SLSNe) have been documented to  exhibit  an early peak that can be as luminous as that of a normal SN before  slowly rising to maximum brightness months later \citep[e.g.,][]{nicholl16}.

 A small number of stripped-envelope SNe without hydrogen (Type~Ib) and without helium (Type~Ic) have also been observed at very early epochs with light curve excesses.  
 Among them, the Type~Ib/c SN~1999ex  was the first   with  an early time cooling tail \citep{stritzinger02}, followed by the Type~Ib SN~2008D that was discovered in the moments after explosion through the serendipitous X-ray detection \citep{soderberg08,modjaz09,malesani09}.  Recently, iPTF13bvn is another example of a SN~Ib with an early peak  \citep{fremling16}. 
  
A handful of SNe Ic-BL such as SN~2006aj \citep[e.g.,][]{sollerman06}, GRB-SN~2010bh \citep{cano11}, and SN~2013dk \citep{delia15} were also observed with  double peaks. 
These studies were followed  with the discovery of the  SN~Ic~2013ge that  showed an early peak in the ultraviolet (UV)  \citep{drout16}, and the SN~Ic iPTF15dtg that was  the first spectroscopically normal SN~Ic with an early peak  \citep{taddia16}. 
More recently,  the ultra-stripped SN~Ic iPTF14gqr was found to exhibit  an early peak, perhaps related to     shock-cooling emission from an extended helium envelope \citep{de18}.
Even more recently, ZTF18aalrxas was discovered just after  explosion and reveals  a rise to its first maximum  \citep{fremling19}.  Finally, \citet{xiang19} presented observations of the Type~Ic SN~2017ein that also  show an early peak.

The nature of SE SN progenitor stars  is still a matter of open debate. Many studies suggest a majority of relatively low-mass stars in binary systems \citep{lyman16,taddia18csp,prentice19}, stripped by their companions, while a smaller fraction may originate from massive, compact Wolf-Rayet stars.
Inferring the SN precursor radius and the source of the early emission offers a way to directly identify the exact type of progenitor stars.

In this paper we present early-phase observations of LSQ13abf.  Discovered very close to explosion and classified as a SN~Ib, our observations of LSQ13abf document an early peak. Comparison with  objects  in the literature indicate LSQ13abf is  similar to  SN~2008D, but  also reveals  significant differences such as being brighter in the days following explosion. 

The structure of the paper is as follows: In Sect.~\ref{sec:obs} we describe observations and data reduction; in Sect.~\ref{sec:hg} we present  details regarding the host galaxy, as well as  basic  information on the SN including its distance, metallicity and extinction. Section~\ref{sec:lc} includes the analysis of the SN light curves and broad-band colors and Sect.~\ref{sec:spec} contains our   spectroscopic analysis. In Sect.~\ref{sec:bolo}  the bolometric properties of LSQ13abf are studied, and in Sect.~\ref{sec:model} we model the  bolometric light curve  with semi-analytic models in order to estimate key explosion parameters. The main results are discussed in Sect.~\ref{sec:discussion} and our conclusions are provided in Sect.~\ref{sec:conclusions}.

\section{Data acquisition and reduction}
\label{sec:obs}

The \textit{Carnegie Supernova Project-II} (hereafter CSP-II; \citealt{phillips19}) obtained 20 epochs  of optical $BVri$-band  photometry and a single epoch of near-IR (NIR) $YJH$-band photometry of LSQ13abf. Optical images were  obtained with the Henrietta Swope 1.0 m telescope equipped with a ($+$SITe3) direct CCD camera while the NIR images were taken with the du Pont telescope equipped with the RetroCam imager \citep{hamuy06}.
A complete description of how the science images were reduced is presented by  \citet{krisciunas17}.

Point-spread function (PSF) photometry of the SN was computed and calibrated relative to a local sequence of stars in the field of LSQ13abf. The local sequence itself was calibrated relative to \citet{landolt92} ($BV$) and \citet{smith02} ($ri$) standard-star fields observed over multiple photometric nights. 
 The NIR $J$- and $H$-band local sequences are calibrated relative to the \citet{persson98} standard stars, while the $Y$-band local sequence was calibrated relative to $Y$-band  magnitudes of \citeauthor{persson98} standards presented in \citet{krisciunas17}. 
 
 Photometry of the local sequences in the \textit{standard} system are listed in  Table~\ref{tab:lsq13abf_opt_locseq} (optical) and  Table~\ref{tab:lsq13abf_nir_locseq} (NIR),  respectively, while photometry of LSQ13abf in the CSP-II \textit{natural} system is listed in Table~\ref{tab:optphot} (optical) and Table~\ref{tab:nirphot} (NIR). We note that the local sequence of  stars used to calibrate the NIR photometry have not yet been calibrated relative to standard field observations observed over a minimum of three photometric nights, and we lack NIR template images. In short, the NIR photometry reported for LSQ13abf should be considered preliminary, however, this has no impact on any of the results presented in this paper.

In addition to the CSP-II photometry, we also make use of the early LSQ  photometry computed from images obtained over 5 epochs. These were reduced using an 
IRAF pipeline developed by the LSQ survey \citep{walker15}.
 A host-galaxy template image was  produced from stacking several images of the field obtained prior to the SN discovery and used to subtract the background in each science image using \texttt{HOTPANTS} \citep{becker15}. Photometry of the SN in the LSQ discovery and follow-up images
was computed relative to a local sequence of stars in the SN field,
in an arbitrary natural $gr$-band system (arbitrary zero point)
and then translated into the CSP-II natural $V$-band photometric system by adding a zero point. This works well as both photometric systems are very similar, i.e., their relative color term is $\approx$ 0 for stars with blue to intermediate colors \citep[see][their Fig. 23 as an example]{contreras18}.
The local sequence is listed in Table~\ref{LSQlocseq} and the LSQ13abf $V$-band photometry is listed in Table~\ref{tab:lsq}.

As part of our followup campaign,  four low-resolution spectra of LSQ13abf were obtained with  the Nordic Optical Telescope (NOT) equipped with the  Alhambra Faint Object Spectrograph and Camera (ALFOSC). These are complemented by a single low-resolution spectrum  procured with the Hobby-Eberly Telescope (HET), equipped with the
LRS (Low Resolution Spectrograph). The first spectrum was obtained +3.5~days (d)\footnote{In Sect.~\ref{sec:model}, we estimate the explosion epoch to have occurred on JD 2456395.80$\pm$0.20. This  value is used to compute the phase of our SN observations.} after our estimated explosion epoch  and the last spectrum  was obtained on +60.4~d.

In addition to the five visual-wavelength spectra, a single NIR spectrum was obtained with the Magellan Baade telescope equipped with FIRE (Folded Port Infrared Echellette; \citealt{simcoe13}). Details on the CSP-II NIR spectroscopy program are presented by \citet{hsiao19}.

Visual-wavelength spectra were reduced in the standard manner following the procedures described in \citet{hamuy06}. This includes bias and flat corrections, wavelength calibration using an arc lamp exposure and flux calibration using a nightly sensitivity function derived from observations of a spectroscopic standard star, and no telluric corrections were applied.
Each visual-wavelength spectrum was scaled to an absolute flux level using the $r$-band photometry and if photometry was not obtained on the same night as the spectrum, its brightness was determined with interpolated photometry. In the case of the last $+$60.4~d spectrum, calibration was done relative to photometry estimated via linear extrapolation.  
The FIRE spectrum was reduced using the {\tt firehose} software package developed by \citet{simcoe13}. The reduction steps are described by \citet{hsiao19}. Details of the optical and near-infrared spectroscopic observations are given in Table~\ref{tab:spectra}.

\section{Host-galaxy properties, reddening and distance}
\label{sec:hg}

\begin{figure}
\includegraphics[width=16cm]{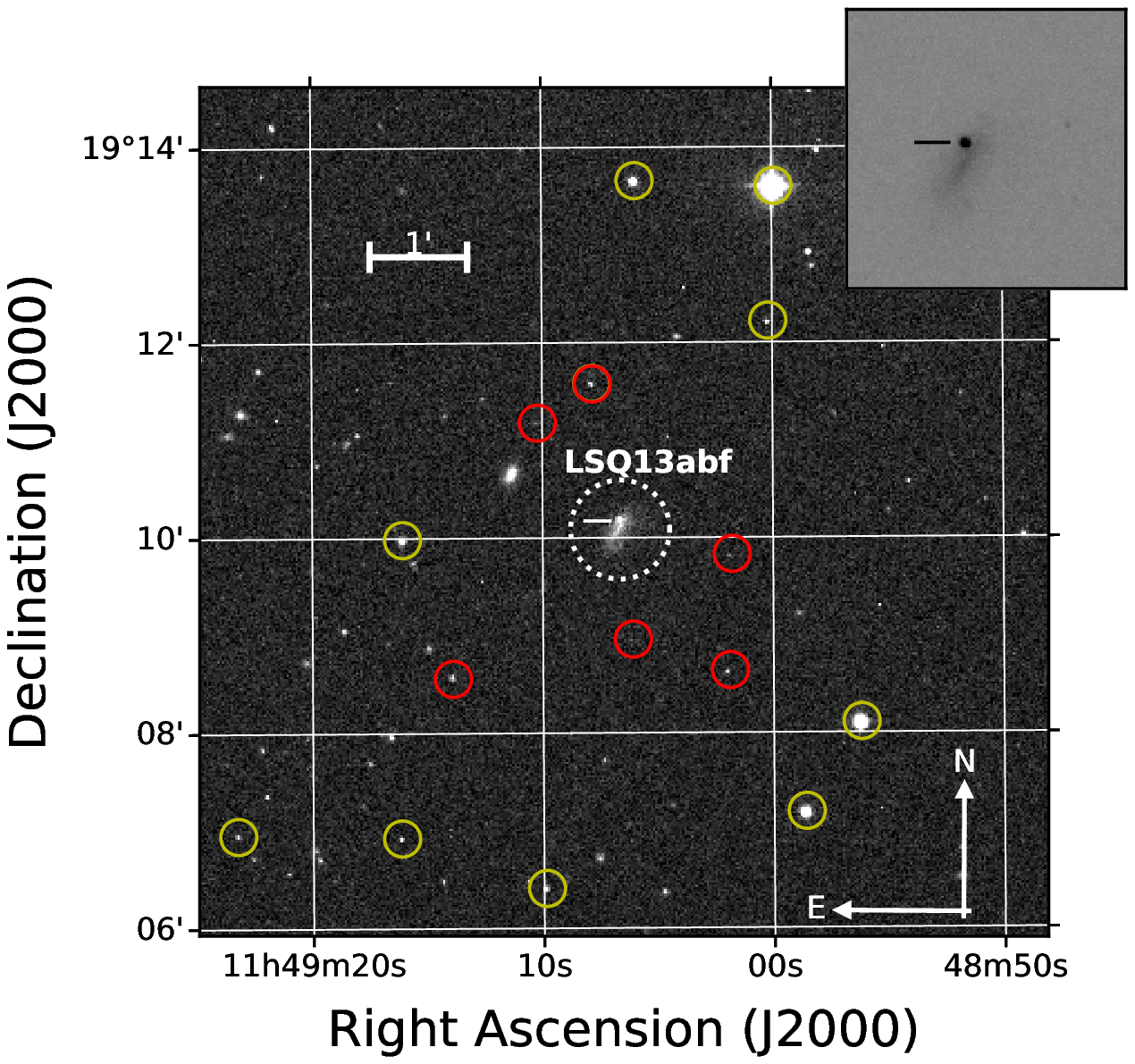}
\caption{\label{FC}Finding chart of LSQ13abf constructed using a single $r$-band image obtained with the Swope 1-m telescope when the supernova was at peak brightness.The position of the SN (dotted circle) is shown within the inset in the upper right corner of the figure, and the  optical (yellow circles) and NIR (red circles) local sequences are indicated.}
\end{figure}

Supernova LSQ13abf was discovered by the LaSilla-QUEST Survey (LSQ; \citealp{lsq}) with an apparent $m_{gr}$-band magnitude of 18.6~mag. 
The previous non-detection of LSQ13abf (with a limiting $gr$-band magnitude of 19 mag) dates 342 observer-frame days prior to discovery, which occurred on JD~2456397.53 (15.03 April 2013 UT). 
The transient was located in  SDSS~J114906.64$+$191006.3 and has coordinates R.A.(J2000.0) $=$ 11h49m06s.62 and Decl.(J2000.0) $= +19^\circ$10$'$10$\farcs$71.
A finding chart with the position of LSQ13abf in SDSS~J114906.64$+$191006.3 is provided in Fig.~\ref{FC}. 
LSQ13abf was initially classified by the CSP-II as a SN~Ic \citep{morrell13}, however as the object evolved, prevalent \ion{He}{i} lines emerged in the spectra (see Sect.~\ref{sec:spec}). 

Following the procedures of \citet{taddia13met} and \citet{taddia15met}, the de-projected distance of LSQ13abf from its nucleus is $\approx$0.33 times the galaxy radius (r$_{25}$), i.e., the supernova is close to the core of its host. Here we used the position angle, t-type, and the major and minor axes of the host galaxy from HyperLeda\footnote{\href{http://leda.univ-lyon1.fr}{http://leda.univ-lyon1.fr}}.
The host-galaxy center was spectroscopically observed by the Sloan Digital Sky Survey (SDSS; DR13 \citealp{albareti17}).
According to NASA/IPAC Extragalactic Database (NED\footnote{\href{http://ned.ipac.caltech.edu}{http://ned.ipac.caltech.edu}}) the \citet{schlafly11} Milky Way (MW) extinction is $A^{MW}_V = 0.087$~mag when assuming a \citet{fitzpatrick99} reddening law characterized by $R_V = 3.1$.

The host-galaxy extinction appears minimal as the visual-wavelength  spectra of LSQ13abf show no  indications of significant \ion{Na}{i}~D absorption or diffuse-interstellar-band (DIB) features (see Sect.~\ref{sec:spec}). 
Furthermore, since LSQ13abf is a stripped-envelope SN, we can compare its colors between 0 and 20 days past peak to those of the CSP-I unreddened sample of SNe~Ib \citep{stritzinger18b}, and infer a color excess. This exercise is presented in Sect.~\ref{sec:lc} and suggests LSQ13abf is minimally reddened.  
In the following, by extinction correction we explicitly mean Milky-Way reddening correction.

\begin{figure*}
\includegraphics[width=16cm]{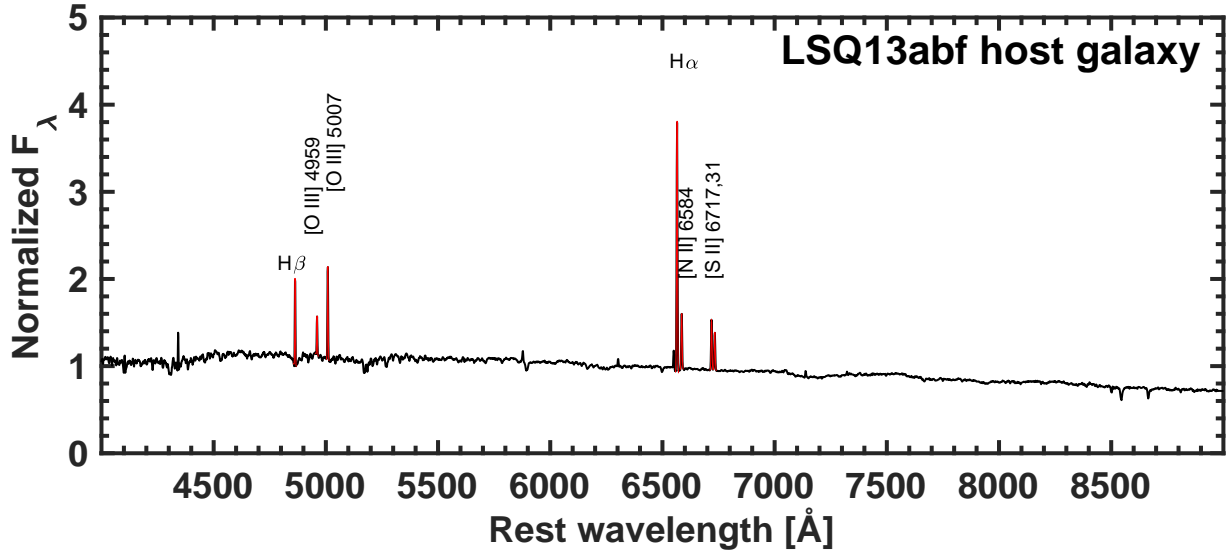}
\caption{\label{hostspec}Host-galaxy spectrum of LSQ13abf. We indicate the main emission lines, which were fit with Gaussian functions (red) to measure the host-galaxy metallicity. The spectrum was corrected for MW extinction. The line fluxes are reported in Table~\ref{tab:host_line_fluxes}.}
\end{figure*}

After correcting the SDSS host-galaxy spectrum for MW extinction, we fit the emission lines of interest (H$\beta$, [\ion{O}{iii}]~$\lambda$5007, H$\alpha$, [\ion{N}{ii}]~$\lambda$6584) with Gaussian functions, as shown in red in Fig.~\ref{hostspec}, to measure the metallicity of LSQ13abf from its SDSS host-galaxy spectrum. The line fluxes are reported in Table~\ref{tab:host_line_fluxes}. We obtain a N2 (and O3N2) \citep{pettini04} metallicity of 12+log(O/H)$=$8.54(8.52)$\pm$0.18(0.16)~dex. 
Assuming a typical $-$0.47~dex~r$_{25}^{-1}$ metallicity gradient \citep{pilyugin04}, implies  LSQ13abf was located in a slightly sub-solar metallicity environment, i.e., 12+log(O/H)$=8.4\pm0.2$~dex. This is also  lower than the average SN~Ib N2.O3N2 metallicity measurements of  12+log(O/H)$=8.7\pm0.2$~dex, estimated from  24 un-targeted SN~Ib in PISCO (\citealt{galbany18}; see their Fig.~8), 
In other words, 20\% of  SNe~Ib are at a more metal poor location.

NED lists the heliocentric redshift of  SDSS~J114906.64$+$191006.3 to be   $z=0.02080$. Adopting  WMAP 5-year cosmological parameters \citep{komatsu09}: $H_{0}=$ 70.5 km~s$^{-1}$~Mpc$^{-1}$, $\Omega_M=$ 0.27, and  $\Omega_{\Lambda}= 0.73$ , this corresponds to the luminosity distance $D_{L} = 98.0\pm6.8$ Mpc and  distance modulus $\mu = 34.96\pm0.15$~mag. Note that this distance includes corrections for  peculiar motions (Virgo + GA + Shapley).

\section{Broad-band light curves and color curves}
\label{sec:lc}

LSQ13abf was discovered by LSQ in survey images obtained with a combined  $gr$ filter.  Inspection of the first epochs of the LSQ light curve reveals an   unusual plateau phase preceding the rise to the main peak. 
Such behaviour is indicative   that the SN was discovered soon after explosion. This is confirmed by the early decline in the $B$-band light curve.  
In Sect. \ref{sec:model} we simultaneously
fit two models to the bolometric properties (one for the early epochs of luminosity, temperature and radius, and one for the later epochs of the bolometric luminosity) to estimate the explosion epoch. These fits suggests the explosion epoch occurred on JD 2456395.80$\pm$0.20 days.  Details covering  the modelling  of the light curve are provided in Sect. \ref{sec:model}. 
In the following, we assume this as the explosion epoch and all phases provided below are relative to this time and corrected to  rest-frame days.  
We note that if we fit a power-law to the first 4 out of 5 epochs of the $V$-band light curve, the derived explosion epoch would be even closer to the first observation; however, we prefer to adopt an explosion epoch based on a physical model.

\begin{figure*}
\includegraphics[width=18cm]{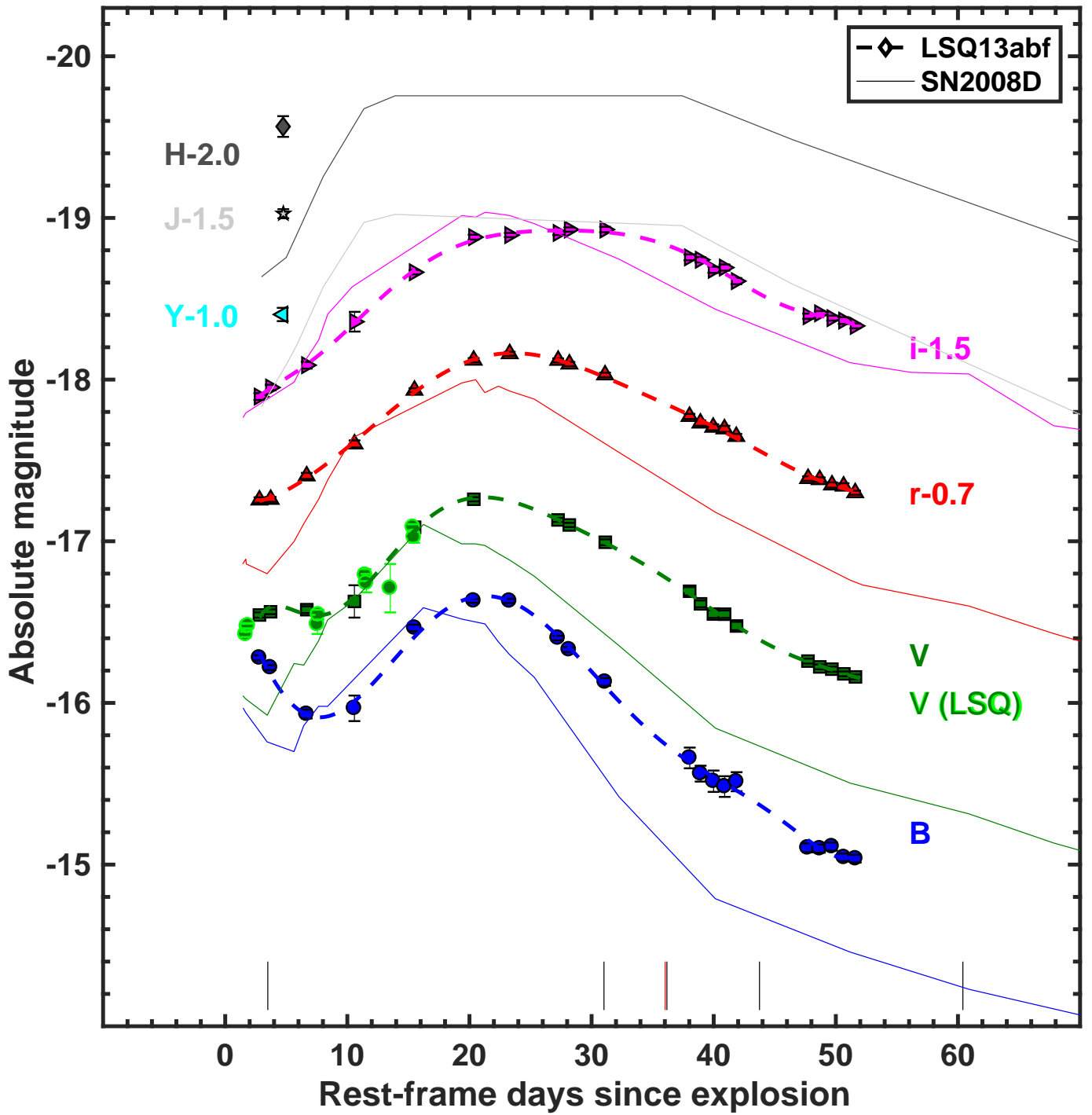}
\caption{\label{lc} Absolute-magnitude light curves of LSQ13abf. Each light curve has been shifted by different amounts in magnitude (reported next to the name of each filter) for visual clarity. The optical light curves were fit with a low-order polynomial marked by dashed, thick lines of the same color of each light curve. The $BVri$ light curves peak at +21.3, +21.1, +23.6, +27.6~d, respectively. The optical(NIR) spectral epochs are marked by black(red) segments at the bottom of the figure. We also report the corresponding absolute magnitudes of SN~2008D \citep{malesani09} as solid colored lines, shifted by the same amounts used for LSQ13abf. The two SNe are remarkably similar when put into context with the light curve diversity among SE SN as shown in Fig.~\ref{abscomplc} of this paper.}
\end{figure*}

Plotted in Fig.~\ref{lc} are the  light curves of LSQ13abf. The photometry has been corrected for reddening and placed on the absolute magnitude scale.
 CSP-II  $BVri$-band imaging  was initiated on  +3~d.
 Over the first week the $B$-band  light curve declines nearly half a magnitude from $-16.3$ mag to $-15.9$ mag. During the same phase, the $V$-band light curve evolution is essentially  flat with an absolute magnitude of $\approx$ $-16.5$ mag.
The $r$- and $i$-band light curves  instead exhibit a steady rise over the same time period. 

A single epoch of  NIR imaging was  obtained after the second optical epoch, close in time to the minimum of the early $B$-band light curve.
  $B$- and  $g$-band photometry obtained on  $+$11~d indicate the SN was  rising steeply to peak brightness, which occurs at $\approx$ +21~d. 
  During the same phase, the $r$- and $i$-band light curves continue to rise to peak, which occur several days later. 
  This behavior is consistent with other stripped-envelope SN samples   (see \citealp{taddia15sdss,taddia18csp}). 
  Upon reaching maximum, the $B$- and $g$- band light curves decline relatively steeply  until +40~d, whereupon  they continue declining but at a  more slower rate. The $r$- and $i$-band light curves follow a somewhat more linear decline after peak. The last epoch of  $Bgri$ photometry was on +52~d. Each optical light curve was fit with a low-order polynomial to provide a continuous representation of the data. From these fits we also obtain robust estimates of the time and magnitude of  maximum brightness in each band, as reported in the caption of Fig.~\ref{lc} and in Table~\ref{tab:peak}.

We now compare our observations of LSQ13abf to those of the well-observed Type~Ib SN~2008D that triggered extensive studies due to its early discovery and the extended followup of  its shock break out and cooling tail \citep{soderberg08,mazzali08,modjaz09,malesani09,bersten13}. 
In Fig.~\ref{lc} the light curves of LSQ13abf are compared to those of the Type~Ib 2008D  presented by \citet{malesani09}. The overall evolution of the two objects  is quite similar with both objects exhibiting an early peak in the bluer bands.  The  $B$-band light curves decline on a similar time scale, followed by a rise to the main peak within a period of several weeks.  The two objects  reach   similar peak magnitudes, though LSQ13abf reaches a slightly brighter peak and has a slightly longer rise time, while post-maximum  SN~2008D  evolves more rapidly. 
Overall, LSQ13abf can be viewed as an analogue of SN~2008D.

\begin{figure*}
\includegraphics[width=18cm]{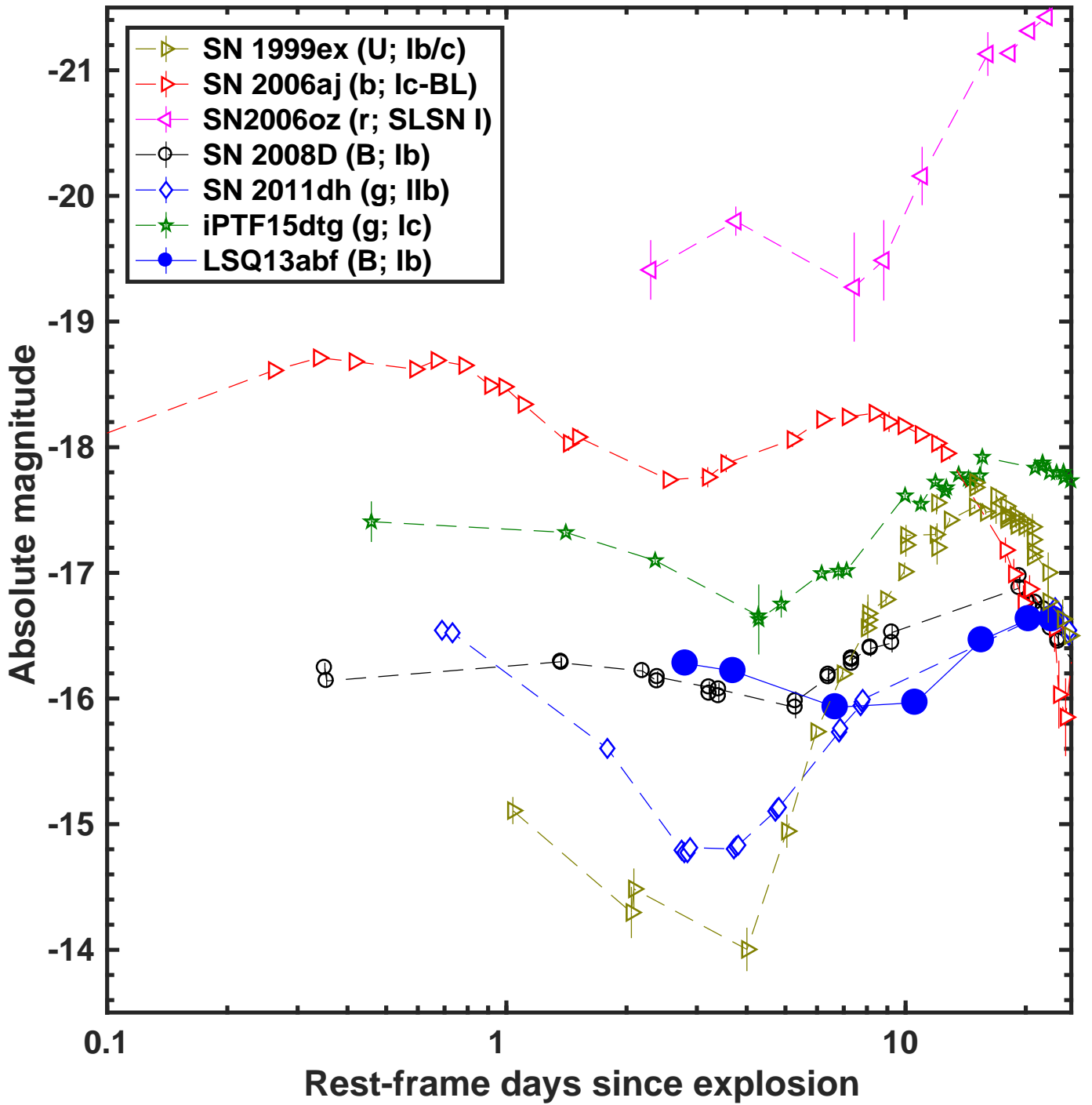}
\caption{\label{early_lc_comp}
Comparison of the early $B$-band  light curve of LSQ13abf to the early  $U,b,B,g$ and $r$ light curves of other SE~SNe (one per sub-type), and for completeness the hydrogent dificient (Type~I) SLSN 2006oz \citep{leloudas12}, which also exhibits a double peak. In the legend we report the filter and the SN type for each SN included in the comparison. The data for SNe 1999ex, 2006aj, 2006oz, 2008D, 2011dh, and iPTF15dtg are taken from \citet{stritzinger02}, \citet{brown09}, \citet{leloudas12}, \citet{bianco14}, \citet{arcavi11}, and \citet{taddia16},  respectively.}
\end{figure*}

We can appreciate how similar their early evolution is by comparing  in Fig.~\ref{early_lc_comp} the early-phase photometry of LSQ13abf and SN~2008D to those of other objects (one per subtype) that also exhibit an early peak in their light curves.  The sample of  stripped-envelope SNe exhibit early peaks ranging from absolute magnitudes between $-$15 mag to $-$20 mag, and reach minima occurring between $+$2.5~d to $+$8~d. There are only a small number of SE SNe that are not SNe~IIb or SLSNe with early light curves excesses in the literature.  As is discussed in Sect.~\ref{sec:intro}, LSQ13abf is the fourth SN~Ib with a documented early peak.

\begin{figure*}
\includegraphics[width=17cm]{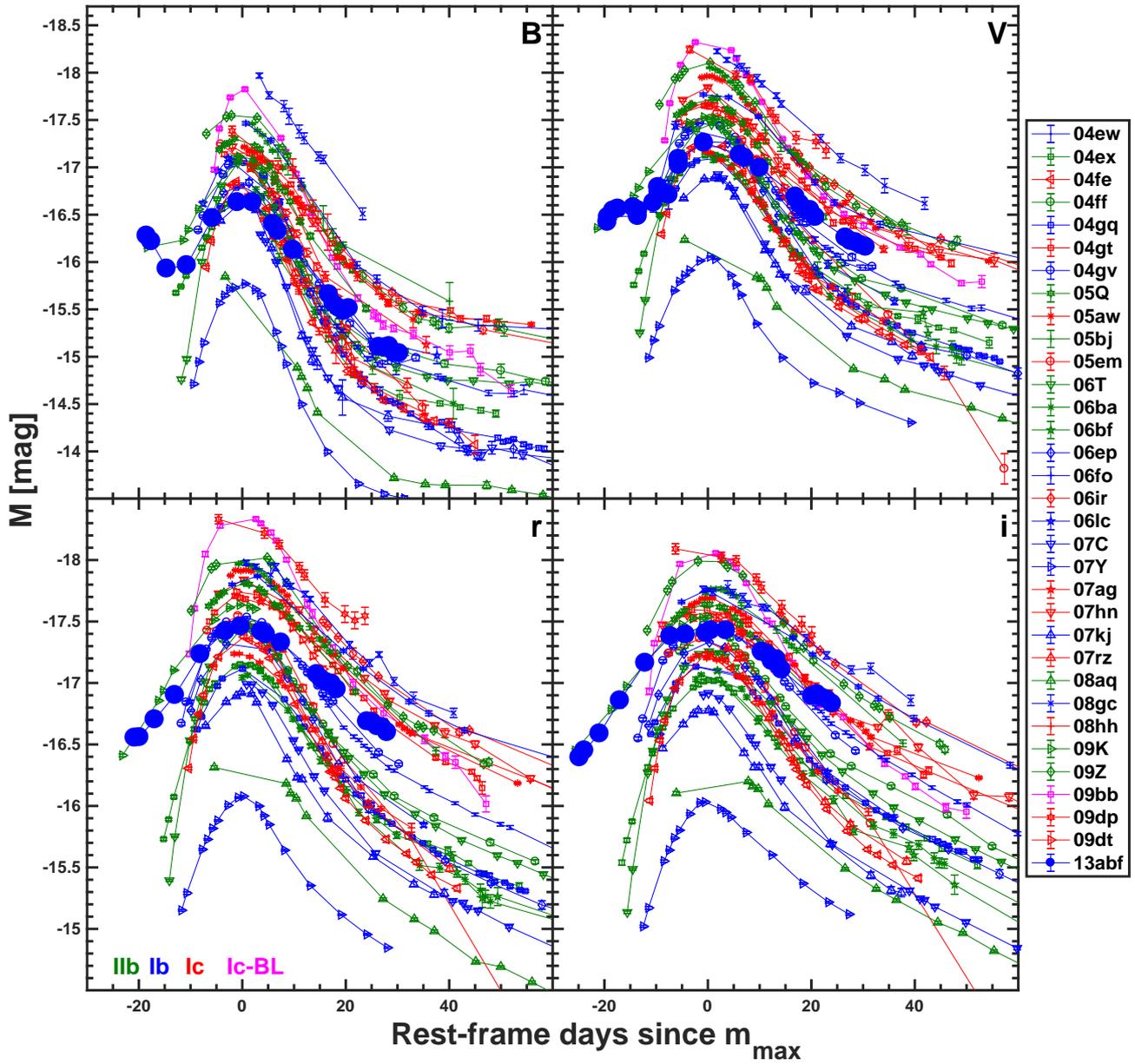}
\caption{\label{abscomplc}Absolute magnitude, $BVri$-band light curves of LSQ13abf compared to those of the CSP-I SE~SN sample \citep{stritzinger18a,taddia18csp}. Objects are color coded relative to spectroscopic sub-type as indicated in the bottom left panel.} 
\end{figure*}

We now turn to comparing the overall broad-band light curves of LSQ13abf to  those of the CSP-I SE~SN sample  \citep{stritzinger18a,taddia18csp}. 
As seen from inspection of  Fig.~\ref{abscomplc}, the absolute $r$-band light curve peaks at $-$17.46$\pm$0.15 mag (the error is dominated by the uncertainty on the distance), placing it near  the middle of the luminosity distribution for the comparison SN~Ib  sample ($-17.22\pm0.60$ mag), 
Interestingly, LSQ13abf is also found to be relatively broader  on both the rise and decline from maximum. Quantitatively,  a  $\Delta m_{15}(r)=0.4\pm0.02$~mag is measured directly from the $r$-band light curve, indicating LSQ13abf evolves nearly a factor of two more slowly in the first two weeks post maximum   compared to the average value of 
$\Delta m_{15}(r)= 0.75\pm0.21$ inferred from the
CSP-I SN~Ib sample \citep[see][]{taddia18csp}.
LSQ13abf takes 23.6$\pm$0.3 days to rise to $r$-band peak.
This is two days longer than the average rise time of  SNe~Ib (21.3$\pm$0.4 days), as inferred from the SDSS-II SN survey \citep{taddia15sdss}.
We  note also that in Fig.~\ref{abscomplc}  the Type~IIb SN~2009K also exhibits an early peak in the bluer bands, resembling both LSQ13abf and several other SNe~IIb in the literature. 
The rising part of LSQ13abf's $r$-band light curve is characterized by a difference in magnitude 
 of $\Delta m_{-10}$~$=$~0.35~mag between peak and 10 days before peak. This is among the lowest values of this parameter measured from the SNe~Ib and SNe~Ic analyzed in \citet{taddia15sdss}. Clearly,  LSQ13abf is characterized by a broad light curve.

\begin{figure*}
\centering
\includegraphics[width=14cm]{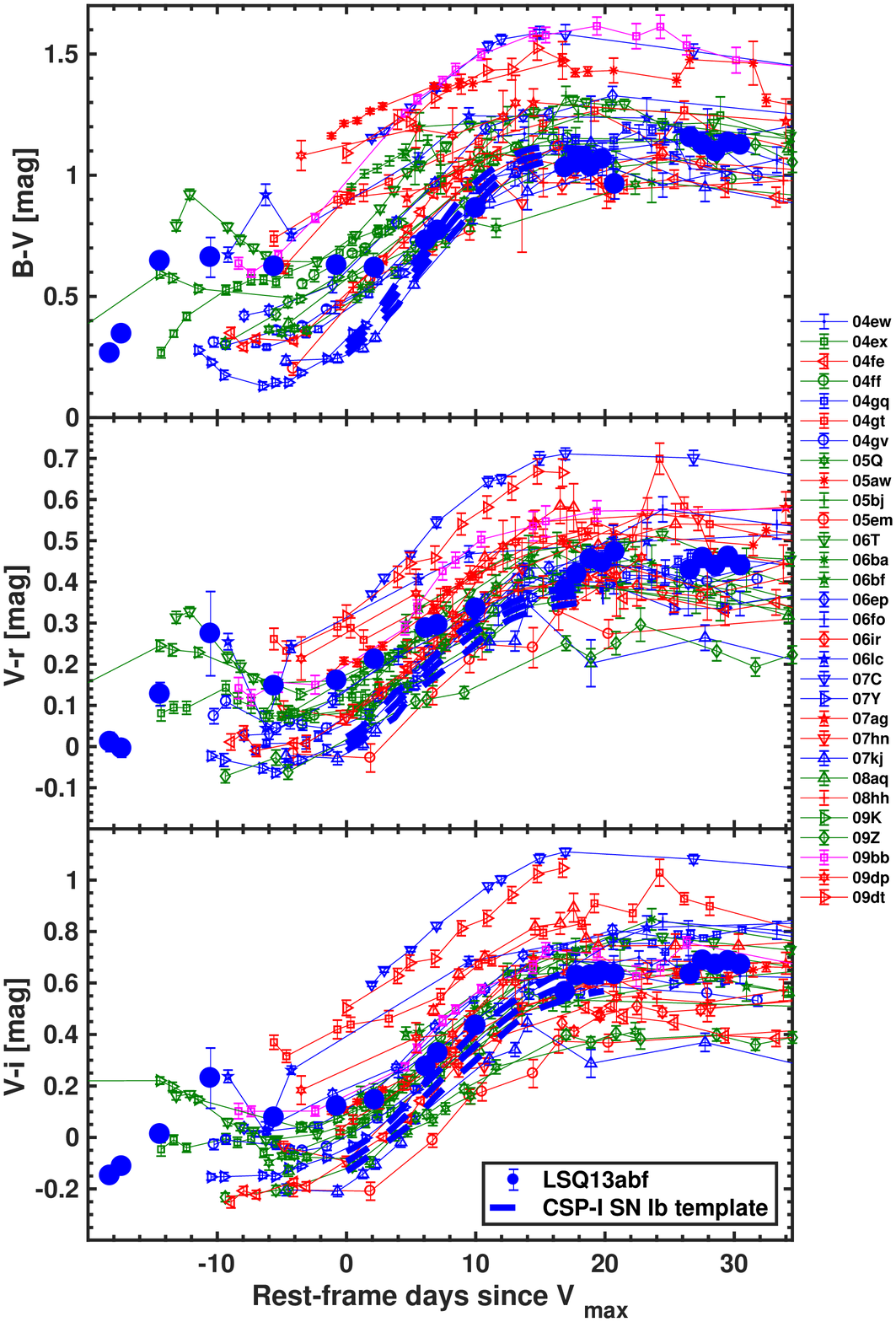}
\caption{\label{color} Optical  ($B-V$, $V-r$, $V-i$) broad-band colors of LSQ13abf compared to those of the CSP-I SE SN sample corrected for Milky Way reddening \citep{stritzinger18a}.  LSQ13abf is among the bluest objects  and is consistent with the SN~Ib intrinsic color-curve  templates (dashed blue lines; \citealt{stritzinger18b}). This indicates LSQ13abf suffers minimal to no host-galaxy reddening. Note that objects are color coded relative to spectroscopic sub-type as defined in the bottom left panel of Fig.~\ref{abscomplc}.} 
\end{figure*}

We conclude this section  with the inspection of  LSQ13abf's broad-band colors. To do so we compare in Fig.~\ref{color} its  $(B-V)$, $(V-r)$, and $(V-i)$ color curves to those of  the CSP-I SE~SN sample corrected for Milky Way reddening.  Also over-plotted in each panel are the intrinsic color-curve templates for  SNe~Ib presented by  \citet{stritzinger18b}.  LSQ13abf lies among the bluest objects between 0 and 20 days after $V$-band maximum. In particular, its $(B-V)$ color curve nearly overlaps with the Type~Ib SN intrinsic color-curve  templates during this period,  indicating LSQ13abf suffers minimal to no host-galaxy reddening.
Finally, we note the rapid color evolution  exhibited by LSQ13abf in the first epochs of followup  corresponds to  the early post-shock breakout cooling phase. The rapid evolution in color  is consistent with observations of SN~1999ex \citep{stritzinger02},  while as energy deposition from radioactivity begins to dominate the color-curves evolve similar to those of the bulk of the  comparison objects. 

\section{Spectroscopy}
\label{sec:spec}

\begin{figure*}
\includegraphics[width=12cm]{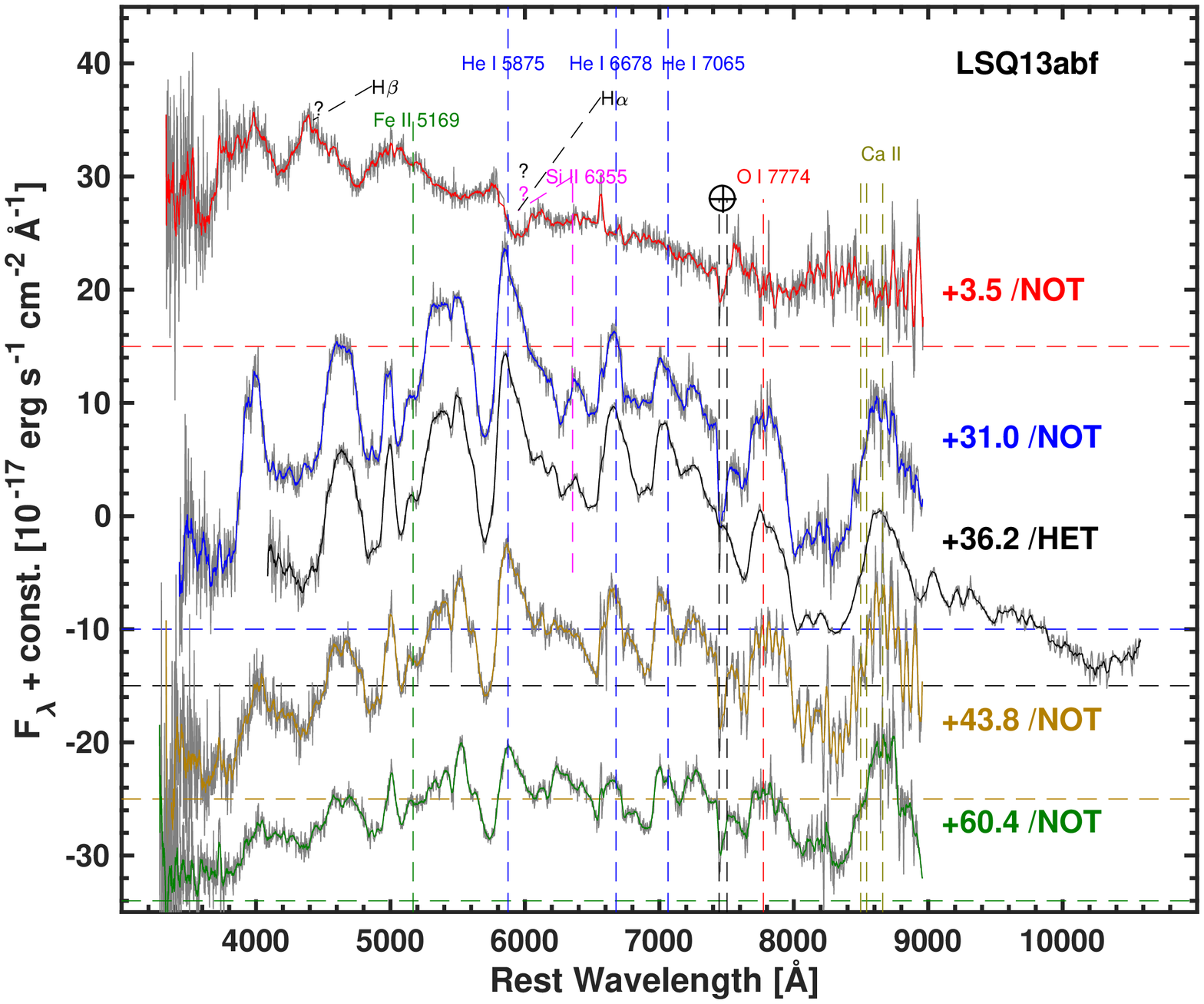}\\
\includegraphics[width=12cm,height=9cm]{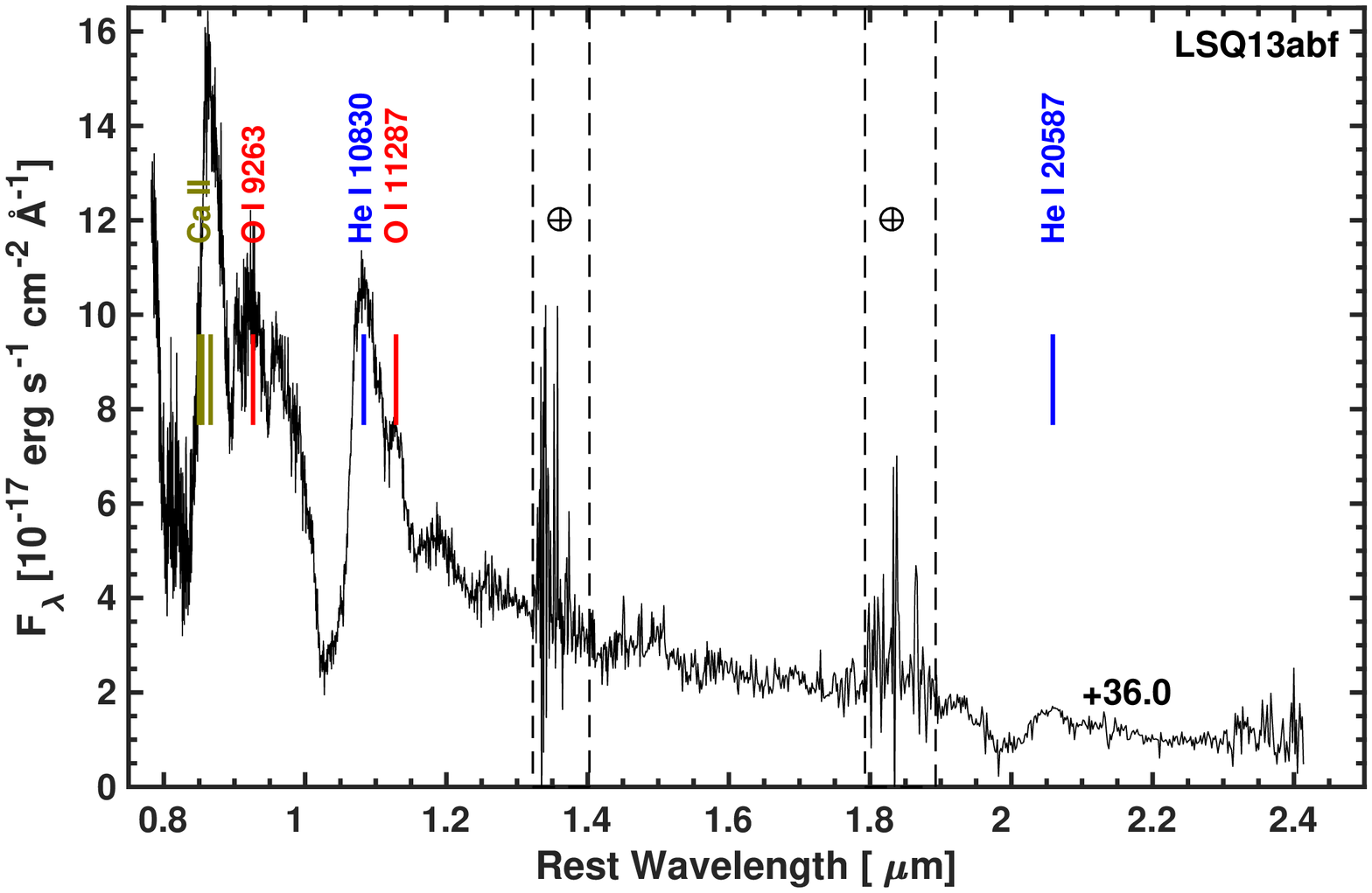}
\caption{\label{spec}\textit{top:} Spectral sequence of LSQ13abf. Each visual-wavelength spectrum was scaled to an absolute flux level using the $r$-band photometry and are shown at rest wavelength. Reported next to each spectrum is its phase (rest-frame days from explosion epoch) and  telescope used to make the observation. The zero-flux level for each spectrum is marked by a horizontal dashed line having the same color of the smoothed spectrum. The main spectral features at their rest wavelength are marked by vertical dashed lines and labeled. For the first spectrum we report uncertain identifications with question marks and dashed lines pointing to the corresponding features. \textit{bottom:} NIR spectrum of LSQ13abf taken 36 days post explosion. Prominent spectral features and telluric regions are marked. The spectrum was absolute-flux calibrated using the coeval optical spectrum.} 
\end{figure*}

The visual-wavelength spectra of LSQ13abf are plotted in  Fig.~\ref{spec} (top panel).
To provide an aid to the visualization,  over-plotted each raw spectrum (grey)  is a smoothed version (colors). 
The main features  characterizing the spectra include  conspicuous  \ion{He}{i} $\lambda\lambda$5876, 6678, and 7065 features, which are clearly visible starting from the second spectrum taken a month post explosion. The first spectrum is largely devoid of features, and most resembles that of an SN~Ic as indicated by the classification telegram \citep{morrell13}.
Additional features in the second spectrum are attributed to the \ion{Ca}{ii} NIR triplet, \ion{O}{i}~$\lambda$7774 and \ion{Fe}{ii}, with \ion{Fe}{ii}~$\lambda$5169 being the strongest \ion{Fe}{ii} feature and is marked in the figure. There is also a  hint of a narrow H$\alpha$  feature attributed to host-galaxy nebular emission.
 
 The first spectrum exhibits a bluer continuum compared to subsequent ones. This is due to the rapid temperature evolution of the photosphere  as the ejecta expand and cool over time.  The deep absorption at about 6000~\AA\ in the first spectrum could be due to \ion{Si}{ii}~
$\lambda$6355 at $\sim$ $19\,000\pm500$~km~s$^{-1}$, as is thought to be the case in other early SN~Ic spectra (see \citealt{taubenberger06}, their Fig.~8). We label this identification with a question mark in the first spectrum in Fig.~\ref{spec}.  
In the second and third spectra, a feature appears present that could be associated with the 
same ion, \ion{Si}{ii}, but at lower velocity. We note that there could also be contribution from \ion{Na}{i}~D together with \ion{Si}{ii}, as seen in SN~2016coi \citep{prentice18}.
However, \citet{parrent16} questioned the identification of \ion{Si}{ii}~
$\lambda$6355 in early SE SN spectra and suggest that hydrogen could give rise to features between 6000 and 6400~\AA. In our case this would imply a H$\alpha$ velocity of about $\sim$ 27\,700$\pm700$ km~s$^{-1}$.
We label this possible identification with another question mark in Fig.~\ref{spec}. There could be a tiny absorption feature related to H$\beta$ at the same velocity of the alleged H$\alpha$ feature that is also marked in the figure. This would resemble what was observed by \citet[][see their Fig.~2]{parrent16} for the Type~Ic SN~1994I at $-$6~d from peak as well as in the Type~Ib SN~2007Y \citep{stritzinger09}.
The first spectrum also contains prevalent absorption features  at 4200~\AA\ and 4750~\AA, which could be related to \ion{Fe}{ii} as in other SNe~Ic \citep{taubenberger06}. 
The absorption at 4200~\AA\ might also be related to other ions, like \ion{C}{III}/\ion{N}{III} \citep{modjaz09}. 
In SN~2016coi \citep{prentice18}, the feature at 4200\AA\ is attributed to a blend of \ion{Mg}{ii}, \ion{O}{ii}, and \ion{Fe}{ii}. The feature at  4750~\AA\ in SN~2016coi is attributed to \ion{Fe}{ii} blended with \ion{Si}{ii} and \ion{Co}{ii}.

Our single NIR spectrum of LSQ13abf is plotted in the bottom panel of Fig.~\ref{spec} with the location of the two strong telluric regions  indicated with a telluric symbol.
The continuum of the spectrum resembles that of a Rayleigh-Jeans black-body (BB) tail and contains
a  prevalent P~Cygni \ion{He}{i}~$\lambda$10830 feature, along with a weaker P~Cygni feature corresponding to \ion{He}{i}~$\lambda$20587. Features likely attributed   to \ion{O}{i}~$\lambda$9263 and \ion{O}{i}~$\lambda$11287 are  also present, while no discernible hydrogen features are detected. 

\begin{figure*}
\includegraphics[width=16cm]{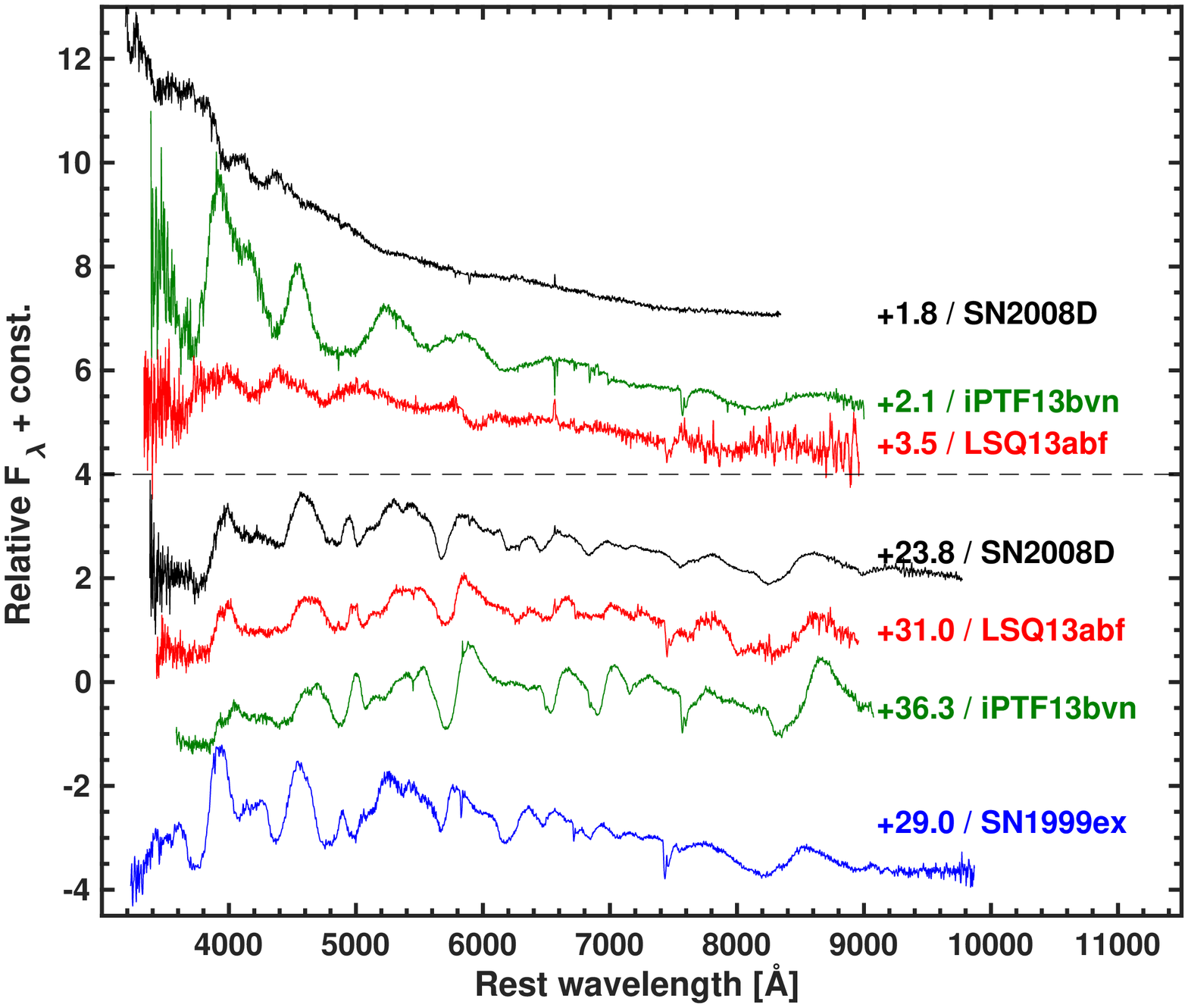}\\
\includegraphics[width=16cm]{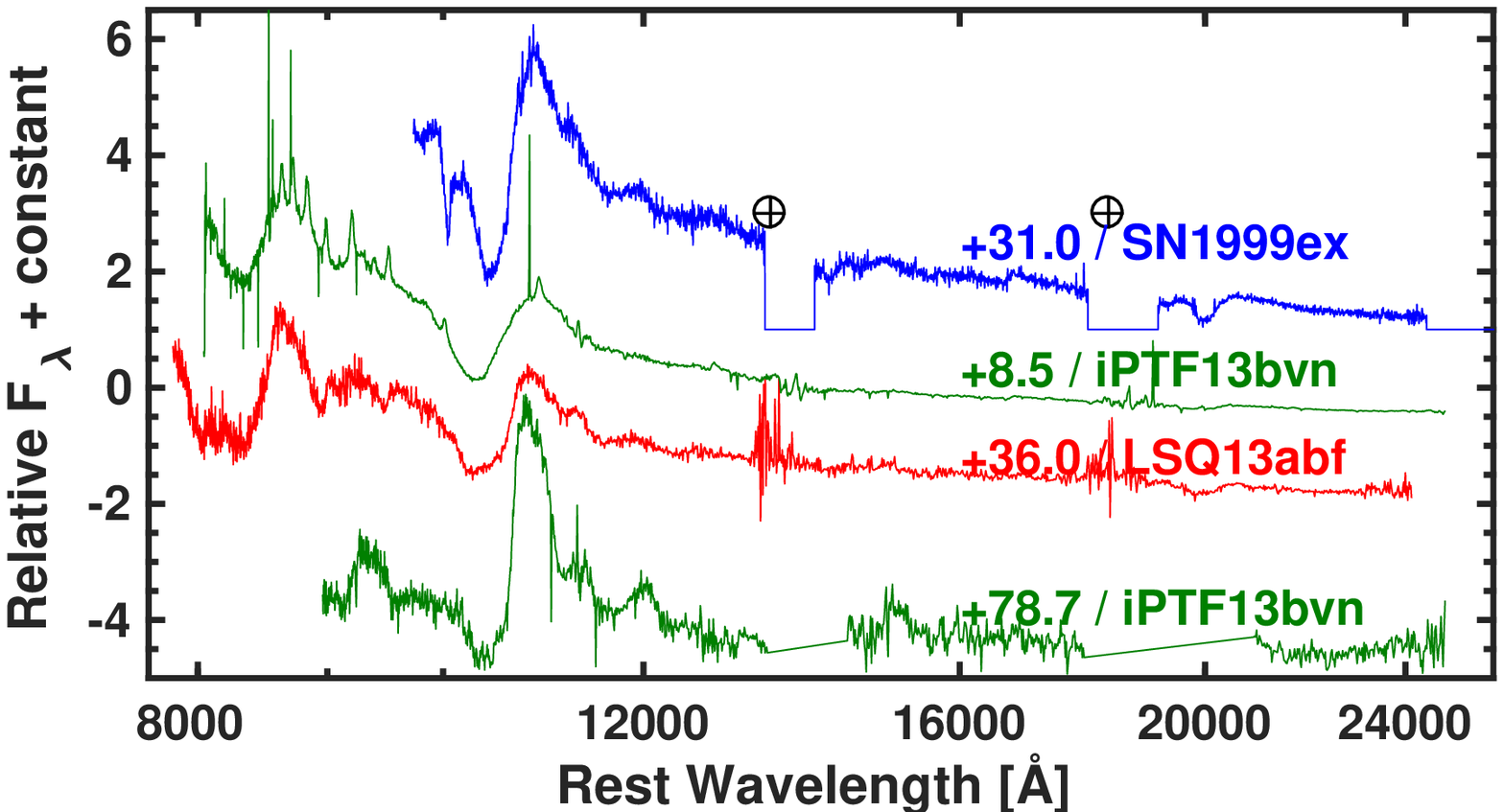}
\caption{\label{speccomp}{\it top:} Visual-wavelength spectral comparison between LSQ13abf and the Type~Ib SN~1999ex \citealp{hamuy02}, SN~2008D \citep{modjaz09}, and iPTF13bvn \citep{fremling16}. {\it bottom:} NIR spectral comparison between LSQ13abf, iPTF13bvn and SN~1999ex.}
\end{figure*}

In the top panel of Fig.~\ref{speccomp}  we compare the visual-wavelength spectra of LSQ13abf to those of  three other well-observed SNe~Ib with cooling tails post explosion. These include SN~1999ex \citep{hamuy02}, SN~2008D (from \citealp{modjaz09}) and iPTF13bvn (from \citealp{fremling16}) at a few days post explosion and a month later. There is a rather large variety of spectral features in the early ``hot" spectra among these objects. iPTF13bvn shows clear \ion{He}{i} features already at $+$2~d. SN~2008D has a spectrum that is basically a continuum with the exception of two clear features below 4500~\AA, associated with \ion{He}{ii} and \ion{N}{iii} by \citet{dessart18} and with \ion{C}{iii}/ \ion{N}{iii} and \ion{O}{iii} by \citet{modjaz09}. These features are also observed in SLSNe \citep{quimby07}. LSQ13abf resembles a SN~Ic, with \ion{Si}{ii} (or H$\alpha$) and two deep absorption features, possibly related to \ion{Fe}{ii} as in SNe~Ic, dominating the continuum. The absorption at 4200~\AA\ is, however, at the same position of the feature identified by \citet{modjaz09} and \citet{dessart18} for SN~2008D and visible in the top spectrum of Fig.~\ref{speccomp}. 
At later epochs, the spectra are almost identical, with similar \ion{He}{}, \ion{Ca}{} and \ion{O}{} features, and therefore LSQ13abf matches a standard SN~Ib. SN~1999ex has weak helium lines and was therefore referred to  by \citet{hamuy02} as an SN~Ib/c. 

 A comparison of our NIR spectrum of LSQ13abf to  NIR spectra of iPTF13bvn at $+$8.5~d  and $+$79~d  and a $+$31~d  spectrum of SN~1999ex is provided in the bottom panel of Fig.~\ref{speccomp}. Overall the spectra   are broadly similar with the \ion{He}{i}~$\lambda$10830 line dominating the spectral region, though there are some differences in the strength and position of the minimum absorption of this feature. 

\begin{figure*}
\includegraphics[width=16cm]{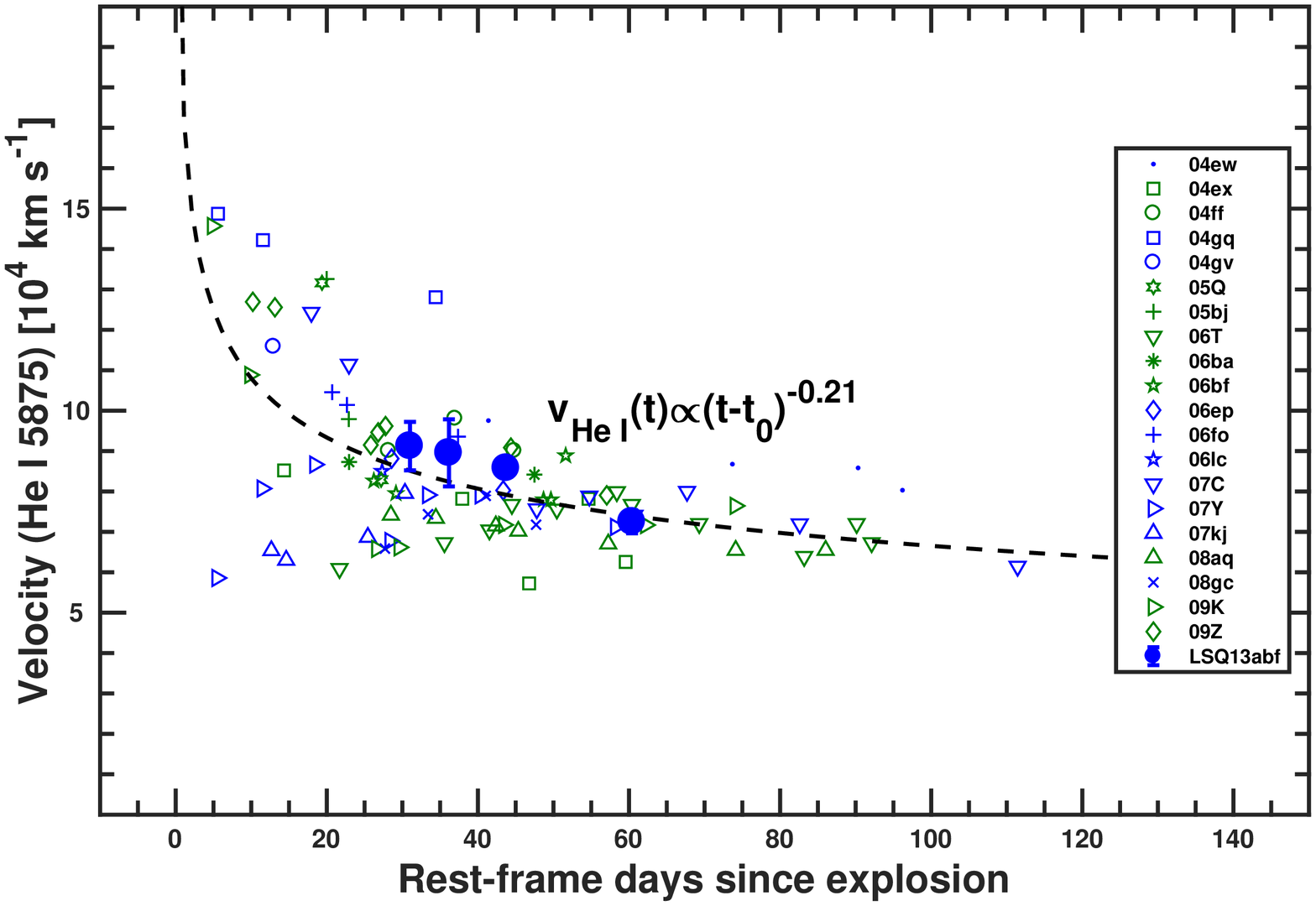}
\includegraphics[width=16cm]{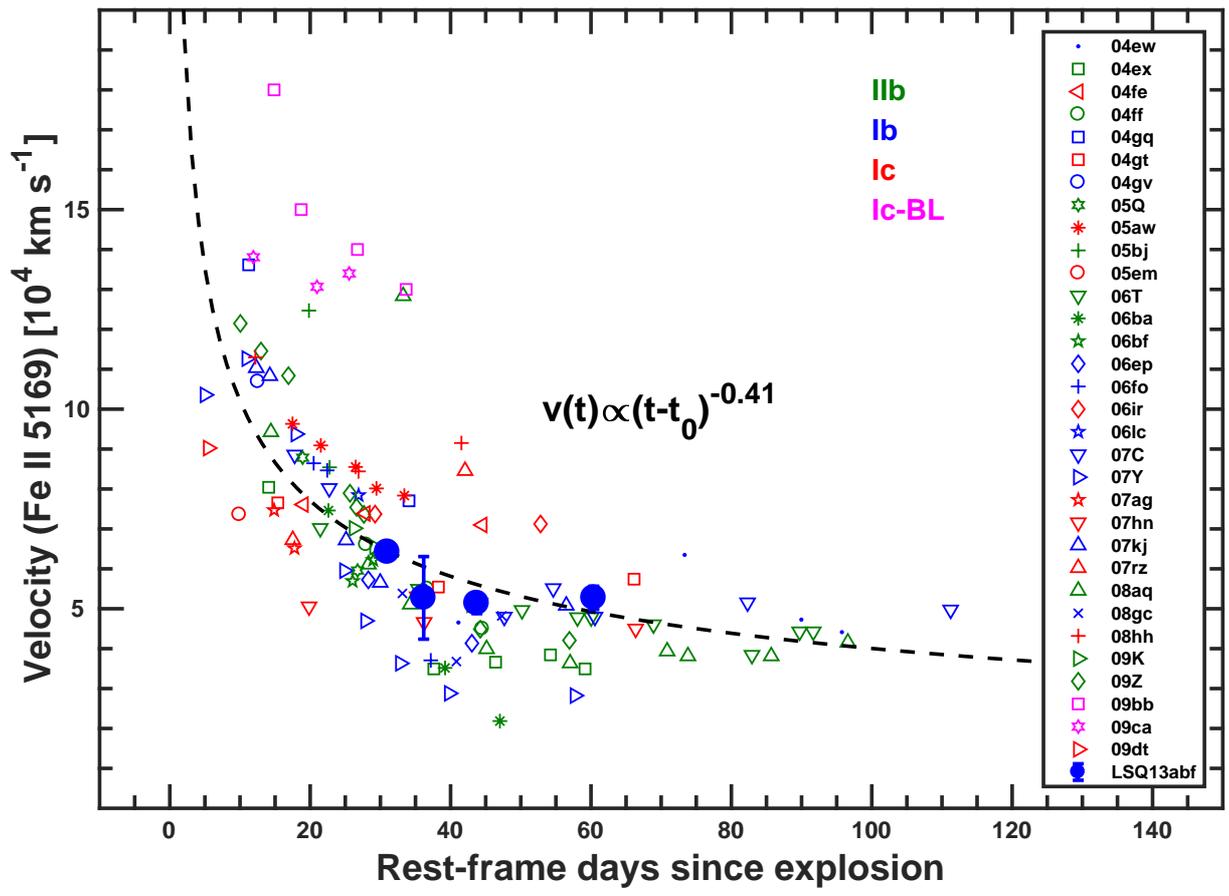}
\caption{\label{velocity} \ion{Fe}{ii}~$\lambda$5169 and \ion{He}{i}~$\lambda$5876 Doppler absorption velocities of LSQ13abf compared with values measured for  objects in the CSP-I SE~SN sample \citep{taddia18csp}.}
\end{figure*}

We now present the line velocities of \ion{He}{i}~$\lambda$5876 and \ion{Fe}{ii}~$\lambda$5169 as estimated from the position of maximum absorption of their P~Cygni profile using the spectral analysis software \texttt{misfits} (Holmbo et al., in preparation). The measured velocities are  plotted  in Fig.~\ref{velocity} and reported  in Table~\ref{tab:vel}.  LSQ13abf has expansion velocities similar to the bulk of the CSP-I SE~SN sample, which are also reported along with their power-law (PL) evolution. The inferred expansion velocities of LSQ13abf are used in concert with fitting other bolometric properties  below to derive pertinent explosion parameters.
We note that there is no \ion{He}{i}~$\lambda$5876 nor a clear \ion{Fe}{ii}~$\lambda$5169 detection in the first spectrum and thus no velocity measurement.

\section{Bolometric light curve}
\label{sec:bolo} 

\begin{figure*}
\includegraphics[width=18cm]{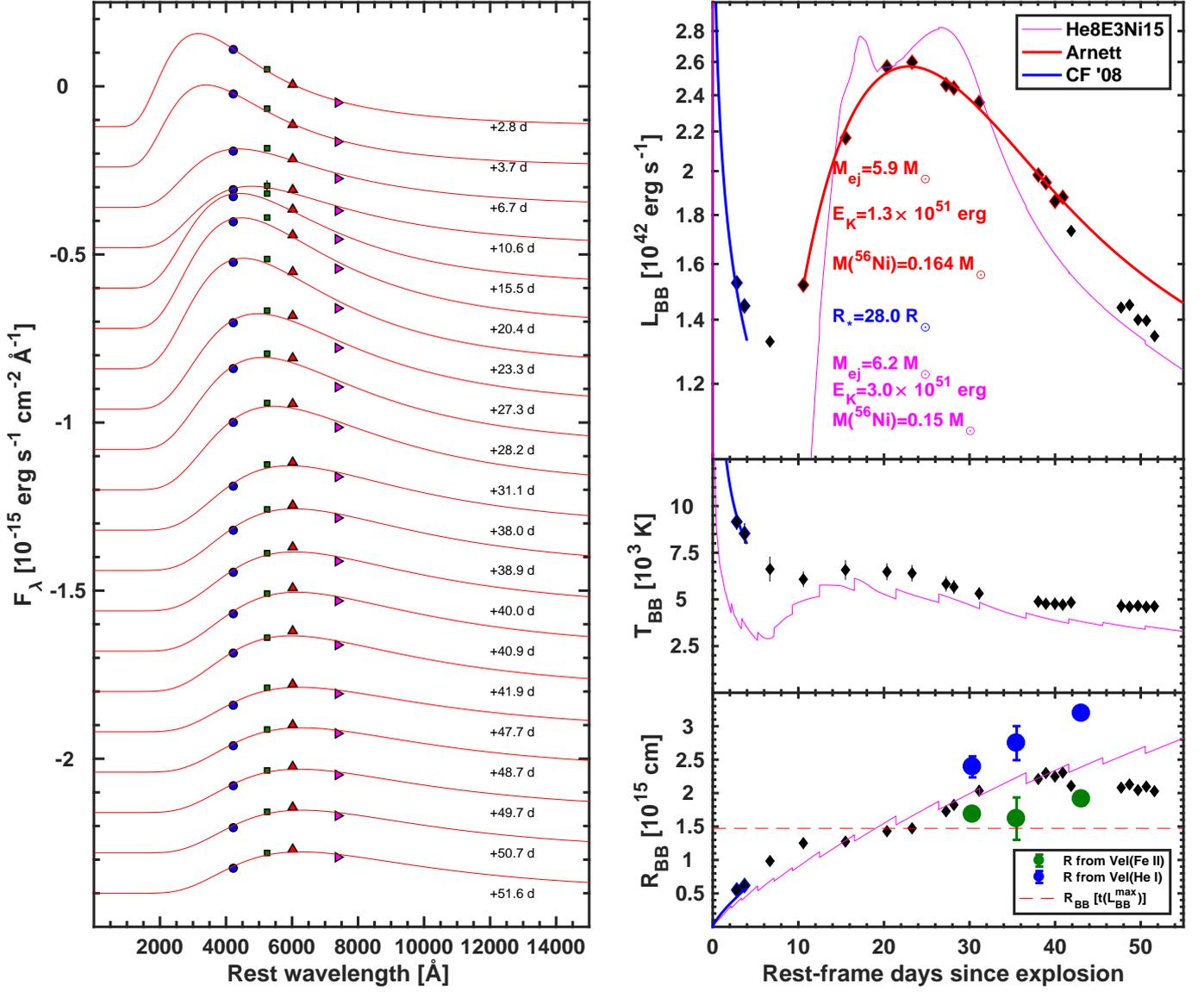}
\caption{\label{bolo} \textit{left:} Spectral energy distributions (SEDs) of LSQ13abf based on  $Bgri$ photometry. The SEDs have been shifted by the addition of an arbitrary constant for clarity and  phases relative to explosion epoch are reported. The symbols for the filters are as in Fig.~\ref{lc}. Each SED is fit with a BB function (red solid line) providing estimates of   (right, top) the total luminosity ($L_{BB}$), (right, middle) the BB temperature ($T_{BB}$), and (right, bottom) the BB radius ($R_{BB}$). 
The early cooling phase is clearly visible with a prompt drop in the $T_{BB}$. \textit{right} In the top panel, $L_{BB}$ is fit with an Arnett model (solid red line), beginning from the fourth epoch post discovery when the process(es) producing the early peak is negligible. To perform the Arnett model fit a peak velocity derived from the  $R_{BB}$ at the time of $L_{max}$  (red dashed line in the bottom panel) was adopted.  This BB velocity at peak was then reduced by $\approx$ 18\%,  which is the   difference between the $R_{BB}$   and the photospheric radius (green dots) derived at later epochs from  the \ion{Fe}{ii} velocity. 
The first two measurements of $L_{BB}$, $T_{BB}$, and $R_{BB}$   were also  simultaneously fit with the post-shock breakout cooling model of \citet[][blue line]{chevalier08}. The corresponding 
 explosion parameters derived from the combined model are reported in the top-panel.  
 Shown in magenta for comparison are the bolometric properties of a hydrodynamical model from \citet{taddia18csp}.}
\end{figure*}

Using our broad-band optical photometry, we proceed to construct spectral energy distributions (SEDs) for each epoch of observation. 
To do so, appropriate AB offsets from Table~16 of \citet{krisciunas17} were first added to the filtered photometry.  Next reddening corrections were applied and the magnitudes were converted to monochromatic flux to construct SEDs, which were then each fit with a BB function. 
This provides a representation of the flux distribution,  including also the wavelength regions  not covered by our data.  The resulting SEDs and best-fit BB functions are  presented in the left-hand panel of Fig.~\ref{bolo}. 
Inspection of the figure indicates that at the earliest epoch the BB functions peak around 3000~\AA. As time evolves the BB peak  progressively shifts towards longer wavelength as the photosphere cools. Cooling follows from the decrease in radioactive energy deposition   combined with  expansion  of the ejecta. The corresponding BB temperature ($T_{BB}$) is plotted in the right-center panel of  Fig.~\ref{bolo}. The early drop in temperature corresponds to the initial decreasing phase in the bluer optical light curves shown in Fig.~\ref{lc}. 
 The value of $T_{BB}$  drops from 9300~K to 6100~K in the first 11 days. Subsequently, $T_{BB}$ remains nearly constant over a fortnight, 
and then again drops, reaching 4600~K by $+$50~d.

By integrating the BB fits over   wavelength from 0 to infinity and multiplying the result by 4$\pi D_L^2$, where $D_L$ is the distance of the SN, we obtain the bolometric light curve of LSQ13abf plotted in the top-right panel of Fig.~\ref{bolo}. The luminosity drops during the first 7 days of evolution, followed by a rise to peak value by $+$23~d. Upon reaching maximum the luminosity  again drops,  reaching the same luminosity of the early minimum (i.e., 1.3$\pm$0.1$\times$10$^{42}$~erg~s$^{-1}$) after $+$50~d. 
LSQ13abf peaks at 2.6$\pm$0.2$\times$10$^{42}$~erg~s$^{-1}$, while our measurement of its first early peak indicates a luminosity of $\approx$ 1.5$\pm$0.1$\times10^{42}$~erg~s$^{-1}$. The errors on these luminosity measurements are dominated by the uncertainty on the distance.

The BB fits also provide a measure of  the BB radius, $R_{BB}$,  which is plotted   in the bottom-right panel of Fig.~\ref{bolo}.  $R_{BB}$ rapidly increases in the first 10 days,  and then the rate-of-increase drops. 
Our BB-fits provide $R_{BB}$   values on the order of 10$^{15}$~cm. 
We overplot the radii computed by multiplying the \ion{He}{i} and \ion{Fe}{ii} velocities shown in Fig.~\ref{velocity} with their spectral phases from explosion time, which are respectively larger and smaller ($\sim$18\%) than the $R_{BB}$. The \ion{Fe}{ii} velocity serves as a good proxy for the bulk-velocity of LSQ13abf's ejecta to be used when modelling the bolometric properties \citep[see also][]{branch02,richardson06}. 

We also compare the bolometric light curve that we obtained through BB fits to that obtained with the SE SN bolometric corrections of \citet{lyman16}, in particular we use their $B$- and $V$-band bolometric correction. The bolometric light curve obtained with the bolometric corrections is similarly shaped, but depending on the epoch it is as much as $\sim$20\% fainter, probably due to the BB encompassing more UV flux than the assumed SEDs in \citet{lyman16}.

\section{Modelling}
\label{sec:model}

\subsection{Early post shock-breakout cooling and Arnett model}

 Armed with the bolometric properties and the velocity measurements, we proceed to infer key explosion parameters of LSQ13abf. The main peak of the bolometric light curve is powered by the radioactive decay of $^{56}$Ni and can be reproduced by an \citet{arnett82} model, which has the $^{56}$Ni mass, the ejecta mass ($M_{ej}$) and the kinetic energy ($E_K$) of the explosion as free parameters. 
 In the model calculations a mean opacity $\kappa_{opt} = 0.07$ cm$^2$~g$^{-1}$  is adopted
 (but see \citealt{wheeler15}), a constant density is assumed for the ejecta, and we impose the condition that $E_K/M_{ej}$ $=$ 3/10 $v_{Fe}^2$ (see \citealp{wheeler15}), where $v_{Fe}$ is the \ion{Fe}{ii} $\lambda5169$ velocity at the epoch of bolometric peak. 
Since there are no $v_{Fe}$ measurements at peak,  but $R_{Fe}$ is about 18\% lower than the value associated with the $R_{BB}$  at later epochs (see the bottom-right panel of Fig.~\ref{bolo}), we use the velocity from the $R_{BB}$  at bolometric peak to infer a value for  $v_{Fe}$. This corresponds to a value of $v_{Fe} = 7300$ km~s$^{-1}$ reduced by 18\%, which is $v_{Fe} =  6000$~km~s$^{-1}$. 

The Arnett model is valid  during the photospheric phase when radioactivity is powering the light curve and there is minimal $\gamma$-ray escape. 
At early epochs, when the first peak is observed and the temperature drops quickly, 
the SN emission is likely produced by another energy source.
Therefore, we apply an alternative model at these early epochs. 
In the literature, the early-phase bolometric properties of SN~2008D were nicely reproduced by an analytic model
describing the cooling of the ejecta after the shock breakout (SBO) \citep{chevalier08}.  
\citet{modjaz09} show a good fit  of this model to  estimates of SN~2008D's $L_{BB}$, $R_{BB}$ and $T_{BB}$. The post-SBO cooling of the ejecta is a promising mechanism to explain the early peak of LSQ13abf, given its similarity to SN~2008D, and  we attempt to fit the same model to LSQ13abf. 
The \citet{chevalier08} model  not only has $M_{ej}$ and $E_K$ as free parameters, but also the progenitor radius ($R^{13abf}_{*}$).

A simultaneous fit of the \citet{chevalier08} model to $L_{BB}$, $R_{BB}$ and $T_{BB}$ during the first two epochs  and of the Arnett model to $L_{BB}$ during the photospheric phase (+4~d to +40~d)   provides  the  $^{56}$Ni mass, $E_K$,  $M_{ej}$,  $R^{13abf}_{*}$, and the explosion epoch. The best fit is shown in the right panels of Fig.~\ref{bolo}, where the Arnett model is plotted with a red line and the \citet{chevalier08} model is plotted with a blue line. The  initial decline phase and the later rise to maximum light are approximately fit by this combined model.
The best-fit explosion parameters from the combined model fit are:  $^{56}$Ni content of $0.16\pm0.01$(0.02) $M_{\odot}$, $E_K$ = [$1.27\pm0.04$(0.23)]$\times10^{51}$ ergs,  $M_{ej}=5.94\pm$0.14(1.09)~$M_{\odot}$,   
an explosion epoch of JD~2456395.80$\pm$0.20,  and $R^{13abf}_{*} =  28.0\pm3.3(6.7)$~$R_{\odot}$. 
The errors quoted outside the  parentheses  correspond to the fit uncertainties and those quoted between parentheses are obtained assuming a 18\% uncertainty on the photospheric velocity,  a further 7\% error  due to the uncertainty on the distance, and in the case of the $^{56}$Ni mass estimates, an additional  10\% error to be conservative. 

 \citet{kathami18} recently provided an analytic model for
determining the $^{56}$Ni mass in radioactively-powered SNe. Their Eq.~A.12 applied to LSQ13abf would indicate a larger $^{56}$Ni mass as compared to that from the Arnett model: 0.27~$M_{\odot}$ vs. 0.16~$M_{\odot}$, assuming $\beta=$9/8).

The analytical work of \citet{arnett82}  is based on a number of underlying assumptions that may pose problems for the application to SE SNe and therefore we should take results from this approach with caution. 
For example, the  Arnett scaling relations depend sensitively on the choice of velocity which does have a number of shortcoming \citep[see][for a discussion]{mazzali13}. Furthermore, there is also an inherent uncertainty when adopting a constant mean opacity as 
highlighted by more sophisticated hydrodynamical model results presented by  \citet[][see their Fig.~19]{dessart16}.
 A constant mean opacity cannot be tuned to match the light curves of models with variable opacity. 
However,  we do stress that making use of  both hydrodynamical  and \citet{arnett82} models to estimate explosion parameters of an extended sample of SE SNe does produce rather rather consistent results, as demonstrated in Fig.~24 of \citet{taddia18csp}.  In the case of LSQ13abf, we note that a  comparison between the constant-opacity Arnett model results and those obtained from a  more sophisticated hydrodynamical model presented below, provide similar explosion parameters.

A constant opacity is also assumed in the  \citet{chevalier08} model. 
Indeed, as the authors mention in their work, a requirement of the applicability of the model is a constant opacity, a condition that will break when the gas recombines to the ground state as the temperature drops.
\citet{dessart11} and \citet{piro13} noticed that when the 
recombination temperature is reached the temperature and the luminosity of a SN remain constant resulting in  a plateau-like evolution. 
This temperature should be around 7000~K (0.6 eV), which is reached by LSQ13abf only after the first two photometric epochs while it exhibits  a period of cooling. This is why to compute the  \citet{chevalier08} model fit to the data only  the first two epochs of photometry are used when cooling is ongoing and the recombination temperature has not yet been reached.

\subsection{Extended-envelope and Arnett model}

\begin{figure*}
\includegraphics[width=9cm]{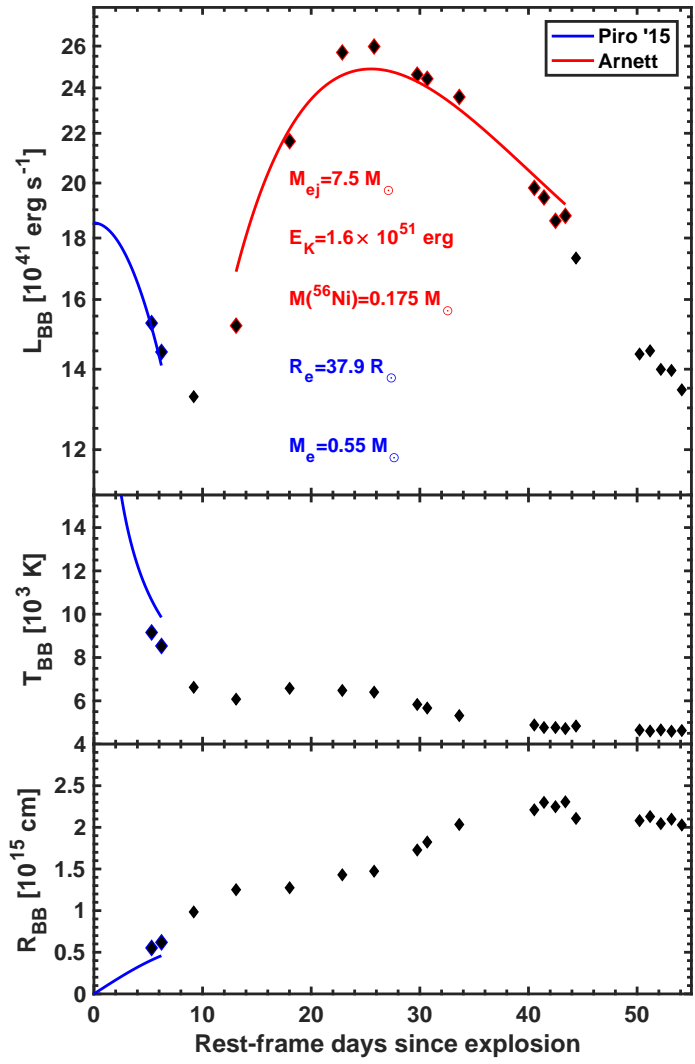}
 \caption{\label{ext_env} As in Fig.~\ref{bolo}, this time the early epochs are fit with a post-shock breakout extended-envelope model \citep{piro15}, and the later epochs are simultaneously fit with an Arnett model. The best fit of this combined model provides a worse match to the early $T_{BB}$ and $R_{BB}$ profiles and to the main peak as compared to the combined Arnett $+$ \citeauthor{chevalier08} model. }
\end{figure*}

We also attempt to reproduce the bolometric properties of LSQ13abf with the combination of an Arnett model for the main peak and the extended-envelope model for the early epochs as in \citet{nakar14} and \citet{piro15}. The result, shown in Fig.~\ref{ext_env}, is worse than that using the model of \citet{chevalier08}. This is not surprising as the early peak is observed in the bluer bands but not in the redder bands, and it was noticed by \citet{nakar14} that this is more consistent with a regular structure of the progenitor star and not with the presence of an extended envelope.
The best fit parameters (see Fig.~\ref{ext_env})  are similar to those obtained by fitting \citet{chevalier08}, again with a progenitor radius of a few $\times$ 10~$R_{\odot}$. In the case of the extended-envelope model the explosion epoch would be 2.6 days earlier than the one estimated by fitting the \citet{chevalier08} model.

\subsection{Companion Interaction and Arnett model}

\begin{figure*}
\includegraphics[width=9cm]{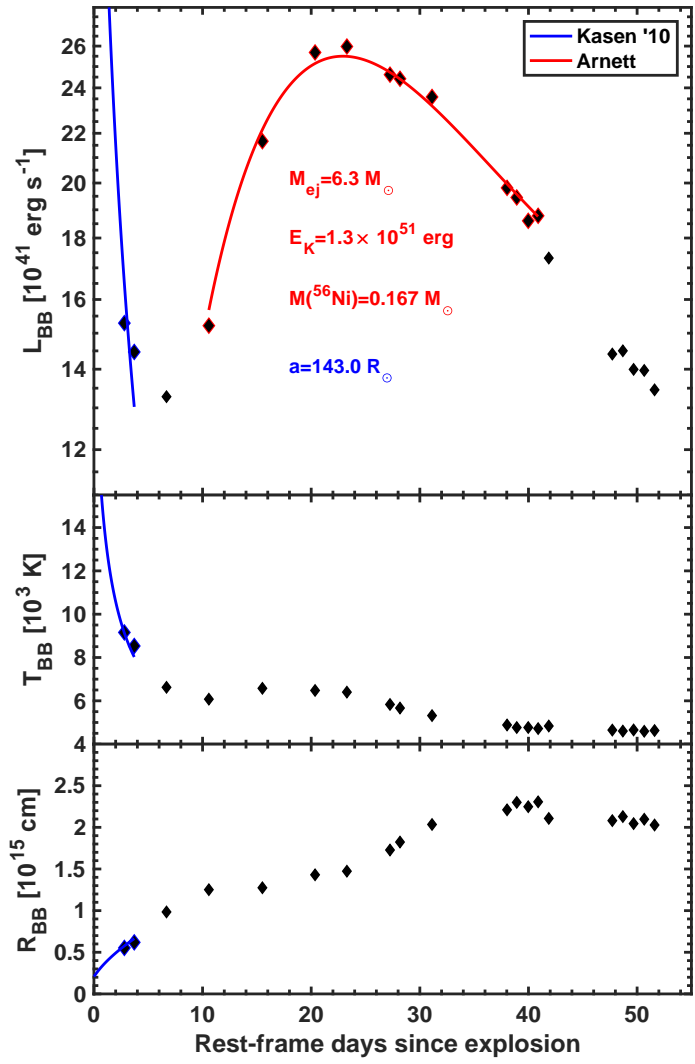}
 \caption{\label{compint} As in Fig.~\ref{bolo}, this time the early epochs are fit with a companion shock-interaction model \citep{kasen10} and the later epochs are simultaneously fit with an Arnett model. The best-fit model reproduces all the observables of the first two epochs. The binary semi-major axis is given by $a=143\ R_\odot.$}
\end{figure*}

A better fit to the bolometric properties at early epochs is obtained by fitting the companion interaction model of \citet{kasen10}, for a binary separation of 143~$R_{\odot}$. The best fit is shown in Fig.~\ref{compint}. Here we assume 45 deg for the viewing angle defined  between the observer direction and the SN-companion interaction region. 
We note that the viewing angle  can significantly  impact the derived binary distance. For example, with a viewing angle of 45 degrees we obtained 143 $R_{\odot}$, but if 10 degrees is assumed, the distance  increases to 2370 $R_{\odot}$.

Despite  the fact that the \citeauthor{kasen10} model produces a good fit to the early light curve of LSQ13abf, population synthesis modeling presented by \citet{moriya15} indicated that the probability of the early light curve of SNe~Ib/c is brightened due to collision is $\approx 0.56\%$. 

\subsection{Magnetar model}

LSQ13abf shows a double peak in the light curves and this makes it different from more typical SNe~Ib observed, for example, in the sample (see Fig.~\ref{abscomplc}). However, LSQ13abf does not resemble extremely peculiar objects like SN~2005bf \citep{anupama05,tominaga05,folatelli06,maeda07} or PTF11mnb \citep{taddia1811mnb}, whose double-peaked light curves had a much longer timescale and for which a magnetar was invoked as one of the possible powering mechanisms. Therefore, in the case of LSQ13abf, there appears to be no evidence for invoking a magnetar model \citep{kasen10_mag}. Even if we assume that the early peak is powered by a magnetar as in \citet{kasen15}, we cannot reproduce the luminosity and the time scale of the early peak given
 the values of $E_K$  and $M_{ej}$ from the previous models of the main peak, for standard values of the magnetic field intensity and the magnetar spin period.
 
\subsection{Hydrodynamical models with a double $^{56}$Ni distribution} 

 We identify a published hydrodynamical model that reproduces the main peak of the bolometric light curves, the temperature evolution, and the radius. This model was among those produced to fit the CSP-I SE~SN sample \citep{taddia18csp} and was computed using a code developed by \citet{bersten11} and  \citet{bersten13}. 
The bolometric light curve of this model --named He8E3Ni15-- is plotted as magenta lines in Fig.~\ref{bolo}. The model  reproduces the observations after $+$10~d when radioactivity dominates the energy deposition and the progenitor radius does not have a significant impact on the shape of the light curve. 

The He8E3Ni15 model is characterized by an $M_{ej} \approx 6.2~M_{\odot}$,  similar to our estimate from Arnett ($M_{ej} \approx 5.94\pm0.14$  $M_{\odot}$). The $^{56}$Ni mass is also similar: 0.15 $M_{\odot}$ vs. 0.16$\pm$0.02~$M_{\odot}$, while the energy is larger in the hydrodynamical model: $E_{K} \approx 3.0\times10^{51}$ ergs vs. 1.27$\pm0.04\times10^{51}$ ergs. The hydrodynamical model reproduces the $R_{BB}$  and not the photospheric radius derived from the $v_{Fe}$  which indicates the need for a larger $E_k$. 
 The progenitor star  used   to produce this hydrodynamical model was a He-rich star that evolved from  a single  star with a  mass of 25~$M_{\odot}$ \citep{taddia18csp}.

\citet{bersten13} presented two different scenarios for  the bright early peak of SN~2008D. This includes  a double $^{56}$Ni distribution and a progenitor structure with an extended envelope, i.e., a star with a dense compact core with a diffuse, extended low-mass envelope. 
In principle, both scenarios   produce a brighter minimum between the early peak and the main peak, as observed in SN~2008D and LSQ13abf. 
 
\begin{figure*}
\includegraphics[width=12cm]{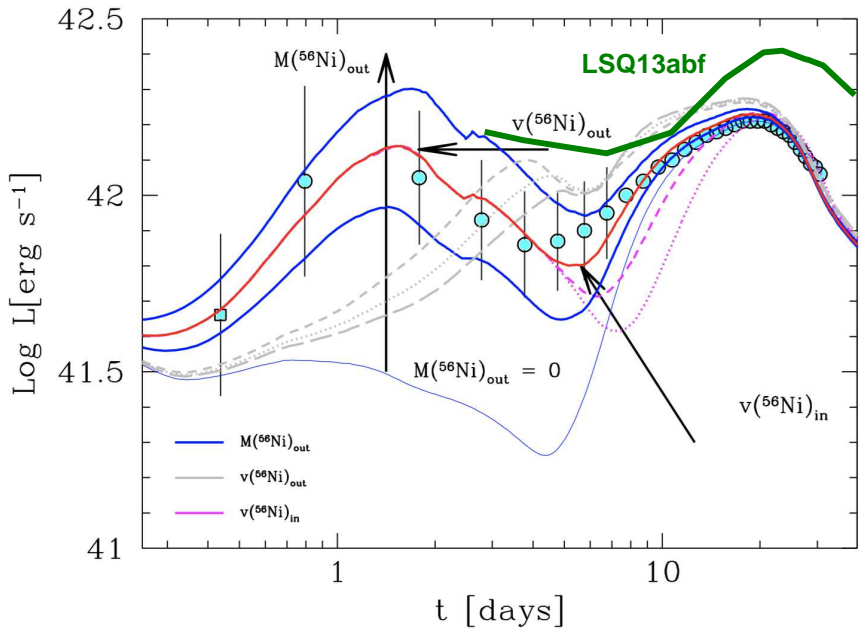}
 \caption{\label{melina_model} Fig.~5 from \citet{bersten13}, adapted to show the comparison between the bolometric light curve of LSQ13abf (green line) to their double $^{56}$Ni-distribution model. The arrows show the effects of increasing the amount of $^{56}$Ni (i.e., M($^{56}$Ni)) in the outer distribution, the velocity of the inner distribution of $^{56}$Ni  (i.e., v($^{56}$Ni)$_{\rm in}$), and the velocity of the outer distribution of $^{56}$Ni (i.e., v($^{56}$Ni)$_{\rm out}$).} 
  \end{figure*}

In  Fig.~\ref{melina_model} the bolometric light curve of LSQ13abf is compared with that of SN~2008D and the \citet{bersten13} double $^{56}$Ni distribution model.
LSQ13abf appears more luminous than SN~2008D in the days and weeks following their explosion epochs. Within the context of a double $^{56}$Ni distribution,  a model with both a larger $^{56}$Ni mass in the inner part and in the outer part and/or  a more centralized inner $^{56}$Ni distribution would allow for a decent fit to the light curve. 

\citet{dessart18} noticed that in the case of SN~2008D, the helium lines are rather narrow around peak. This is in conflict with the presence of  $^{56}$Ni in the other ejecta as suggested by \citet{bersten13}, which would produce broader helium features.

Comparing the spectra of LSQ13abf with those of SN~2008D and iPTF13bvn displayed in the top-panel of Fig.~\ref{speccomp}, reveals LSQ13abf also exhibits narrow helium features with standard expansion velocities (see Fig.~\ref{velocity}).
Given this, the double $^{56}$Ni distribution model is unlikely to provide a better explanation than the post-SBO cooling model.

\begin{figure*}
\includegraphics[width=12cm]{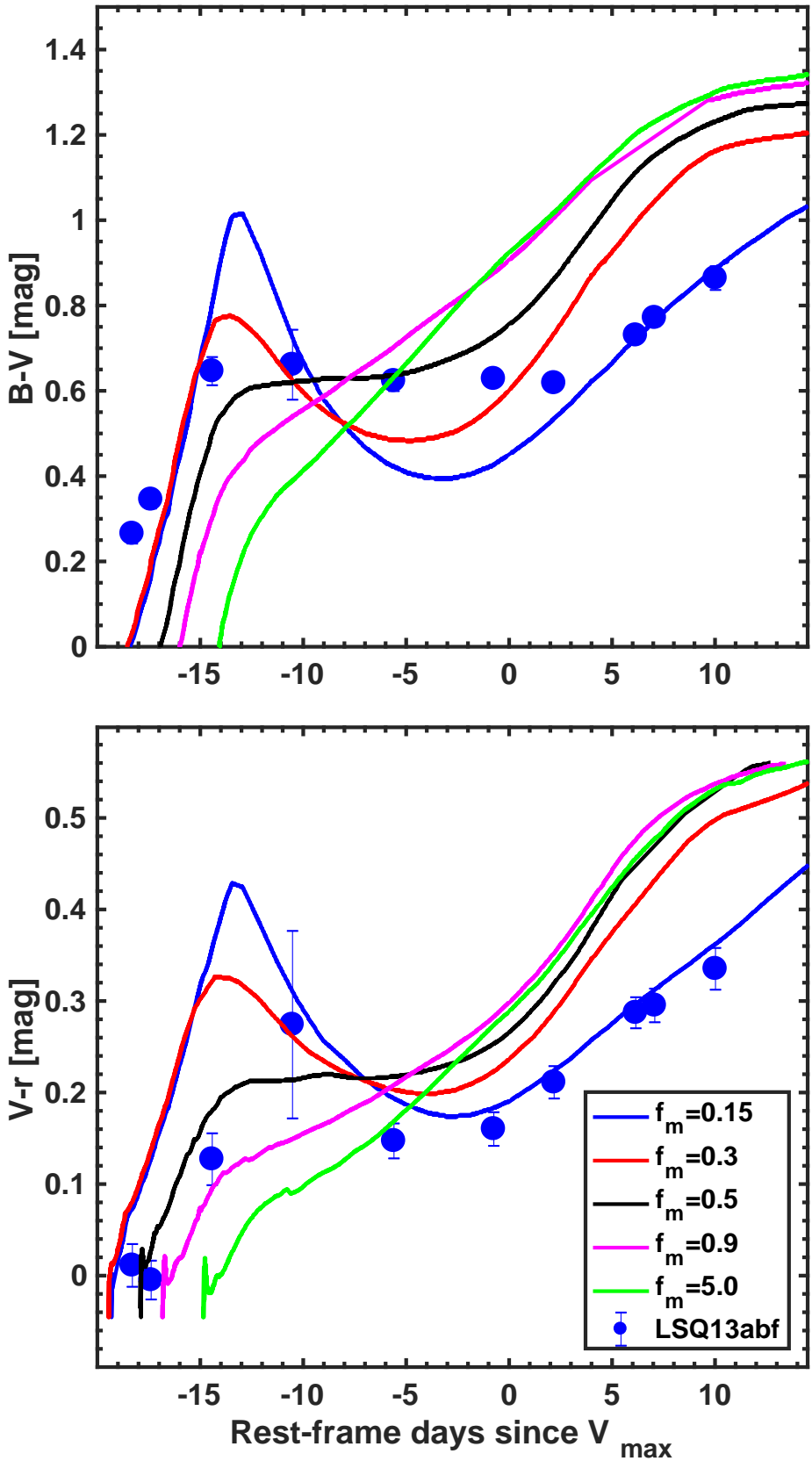}
 \caption{\label{yoon_comp} $(B-V)$ and $(V-R)$ color-curve evolution of LSQ13abf compared to  the synthetric color-curve evolution of five SN~Ib models presented by \citet{yoon19}  covering   various degrees of  $^{56}$Ni mixing ($f_m$).
 The comparison suggests LSQ13abf experienced  a low amount of $^{56}$Ni mixing, i.e., $f_m \sim 0.15-0.3$.} 
  \end{figure*}

A relatively low degree of $^{56}$Ni mixing is also suggested by the comparison of the early colors of LSQ13abf to the models developed by \citet{yoon19}. We show this comparison in Fig.~\ref{yoon_comp}, where low degree of  $^{56}$Ni mixing ($f_m$) produce models with double-peaked color curves, while strong $^{56}$Ni mixing produces monotonically-rising color curves. LSQ13abf shows a double peak in the color curves, and a slow rise to the second main peak, like other SNe~Ib in the literature, and in contrast to most SNe~Ic, as shown by \citet{yoon19} in their Fig.~11.

\subsection{Hydrodynamical models with an extended envelope}

\begin{figure*}
\includegraphics[width=12cm]{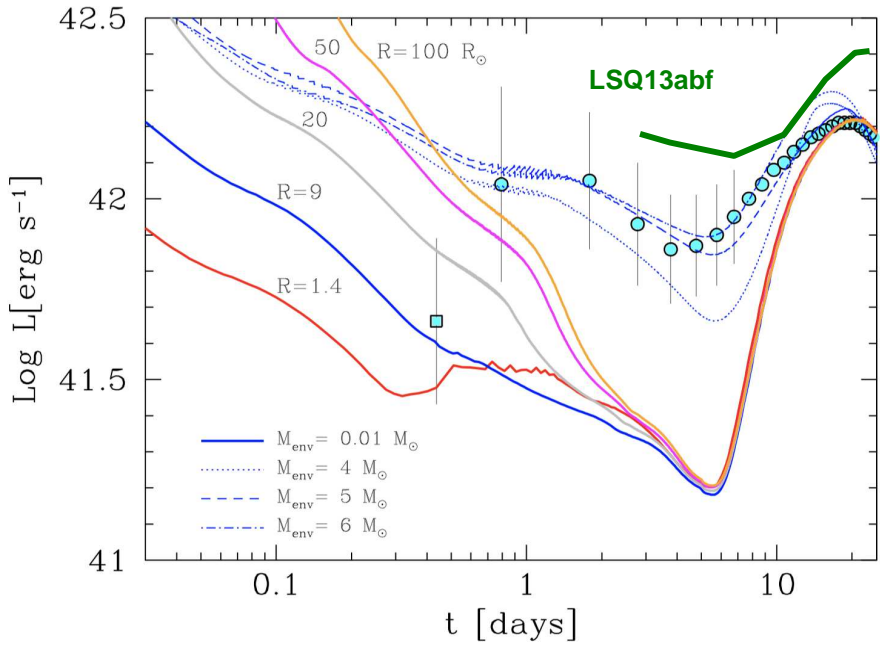}
 \caption{\label{melina_model2}
 Fig. 9 from \citet{bersten13} adapted to show a comparison of the bolometric light curve of LSQ13abf (green line) to their extended envelope models.} 
  \end{figure*}

To facilitate comparison with the findings of \citet{bersten13},  in  Fig.~\ref{melina_model2}   the bolometric light curve of LSQ13abf  is overlaid  on  their  Fig.~9.
The extended envelope models (blue thin lines) do reproduce a bright minimum as in the case of SN~2008D and are characterized by $R^{08D}_{*} \sim 9~R_{\odot}$. LSQ13abf would require a more extended radius due to its  early phase light curve being more luminous,   as well as a more extended structure to explain the luminosity of the early minimum.

\begin{figure*}
\includegraphics[width=9cm]{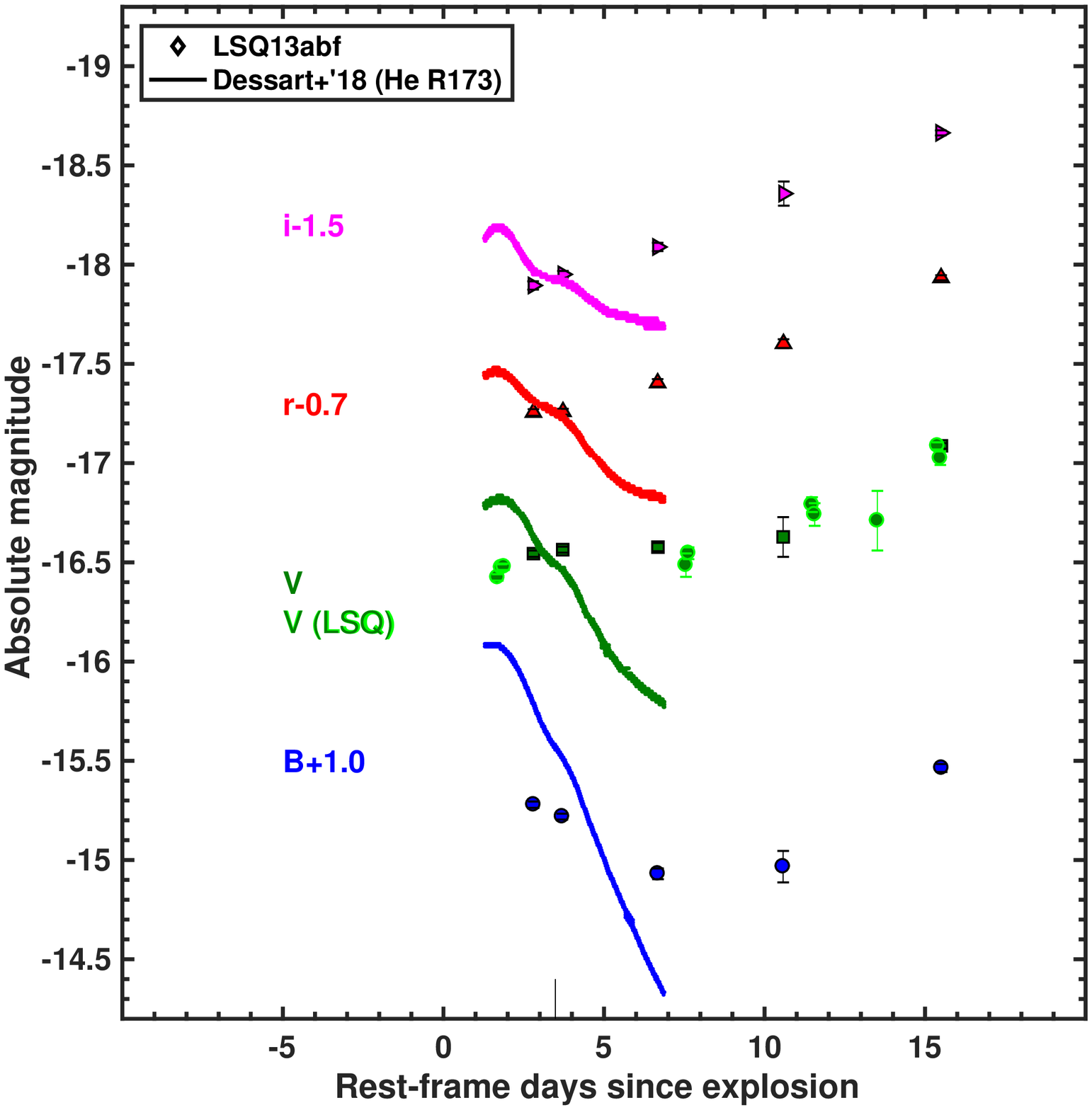}\\
\includegraphics[width=9cm]{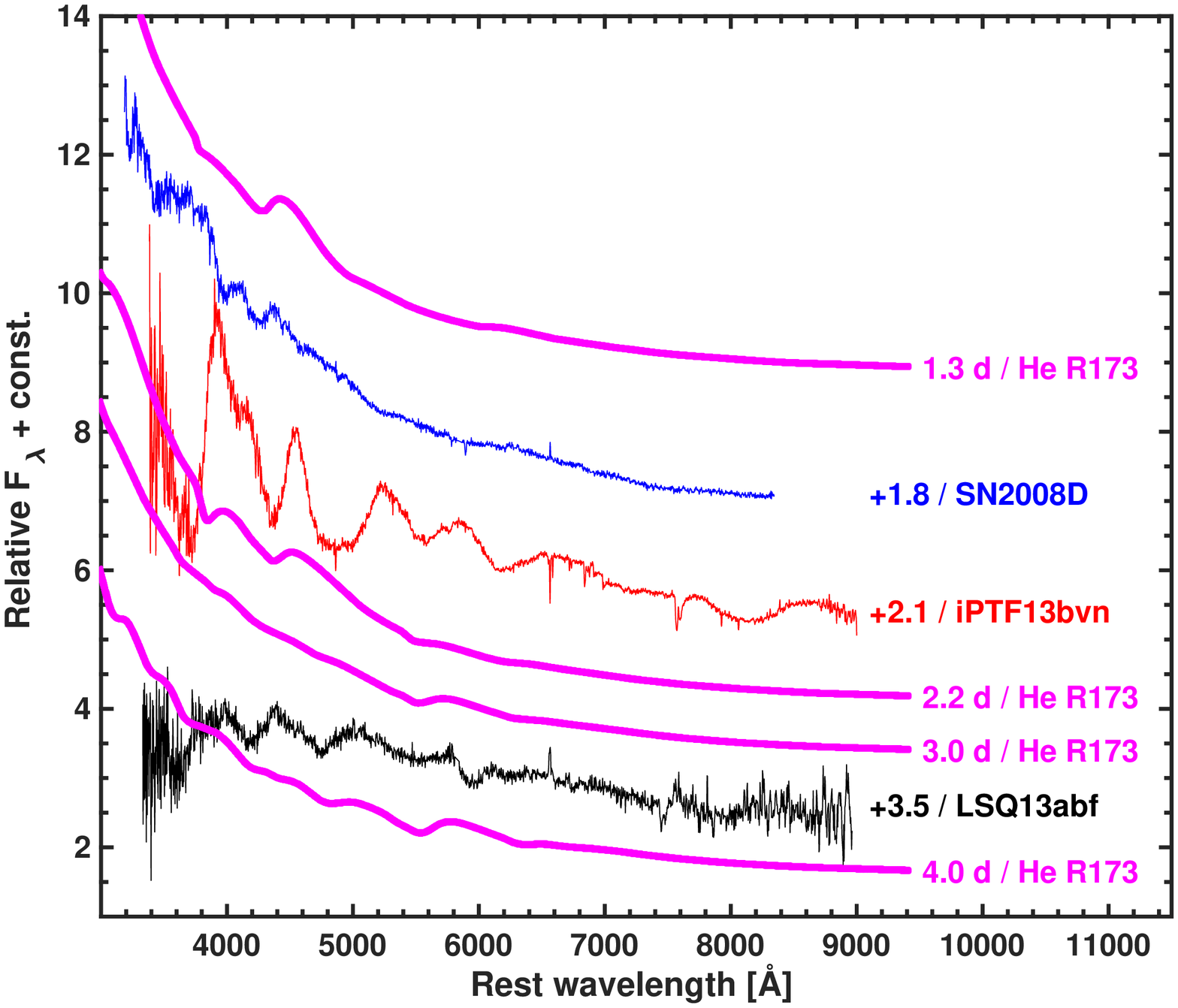}
 \caption{\label{dessart} \textit{top:} Early light curve comparison to the model of the explosion of a He-giant star with extended envelope (173~$R_{\odot}$) from \citet{dessart18}. \textit{bottom:} Comparison of early spectra of the Type~Ib SN~2008D, iPTF13bvn and LSQ13abf with synthetic spectra. }
\end{figure*}

An example of a larger extended envelope is a SN~Ib model presented by \citet{dessart18}. 
Their model produces an early light curve for a He-giant progenitor star with an extended envelope of 173~$R_{\odot}$, a relatively low ZAMS mass (12~$M_{\odot}$), and a pre-SN  mass   of $\sim 2.73$~$M_{\odot}$.  
They account only for the early part of the SN light curve and  do not consider the presence of radioactive $^{56}$Ni. The extended envelope consists of  0.074 $M_{\odot}$,  $M_{ej} 
\sim 1~M_{\odot}$, and  $E_{K} \sim 1\times10^{51}$ ergs. 
Their model overestimates the early luminosity of SN~2008D, however as demonstrated in the top panel of Fig.~\ref{dessart},   the synthetic  absolute $BVRI$-band light curves match reasonably well the first two epochs of the light curves of LSQ13abf. 

The \citet{dessart18}  model has not been fine tuned to match  the main peak of LSQ13abf, as it lacks $^{56}$Ni, and more importantly,  the $M_{ej}$ is too low and therefore would result in a narrower light curve than what is observed. The $E_K$  is similar to our estimate, i.e., $1.0\times10^{51}$ ergs vs. $1.3\times10^{51}$ ergs from our Arnett model. 
A different value of $E_{K}/M_{ej}$ would affect the value for the $R_{*}$ if  the model is fit to the early emission. A higher E$_{K}$/M$_{ej}$ ratio pushes  $R_{*}$ to
 higher  values, assuming the formulas for the extended envelope radius in \citet{nakar14}. Scaling $R_{*}$ in the \citet{dessart18} model using the \citet{nakar14} expression and the values of $E_{K}$ and $M_{ej}$ from our Arnett model, we  obtain an extended envelope radius of 226~$R_{\odot}$  for LSQ13abf. Assuming $E_{K}$ and $M_{ej}$ from the hydrodynamical model shown in Fig.
 \ref{bolo} (magenta lines), then the envelope $R_{*}$ would be 108~$R_{\odot}$. 
In any case, the radius of the extended envelope of LSQ13abf would be larger than that of SN~2008D, i.e., $R^{08D}_{*} \approx 9~R_{\odot}$. 
 In the bottom panel of Fig.~\ref{dessart} the synthetic spectra from the extended-envelope model computed by \citet{dessart18} are compared with  the early spectra shown in the top panel of  Fig.~\ref{speccomp}, including the first spectrum of LSQ13abf and early spectra of SN~2008D and iPTF13bvn.
 The synthetic spectra exhibit \ion{He}{i} lines as soon as \ion{He}{ii}~$\lambda$4686 vanishes (after $+$2~d). There are no discernible helium features in LSQ13abf at this phase. 
 The absorption minimum located in LSQ13abf around 6000~\AA\ is  likely related to \ion{Si}{ii} (or H$\alpha$) and is not present in the model spectra, as well as the strong features at 4400~\AA\ and 4750~\AA, which might be attributed  to \ion{Fe}{ii} features. 
 Our spectrum is not as early as the first SN~2008D spectrum which  according to \citet{dessart18} shows \ion{He}{ii} and \ion{N}{iii} around 4000--4300~\AA. Note however, \citet{modjaz09} attributes these  features to \ion{C}{iii} and \ion{O}{iii}, respectively.

 Despite the differences, the synthetic spectra of \citet{dessart18} for their extended envelope model have diluted features and the \ion{He}{i} features take time to emerge, just as seen  in LSQ13abf.
 A spectral model by \citet[][see their Fig.~8]{dessart18} for a similar progenitor, but without the extended envelope, shows prominent \ion{He}{i} features at early epochs and  a non-diluted spectrum. 

\section{Discussion}
\label{sec:discussion}
\subsection{Model parameters of the early SN~Ib sample}

\begin{figure*}
\includegraphics[width=16cm]{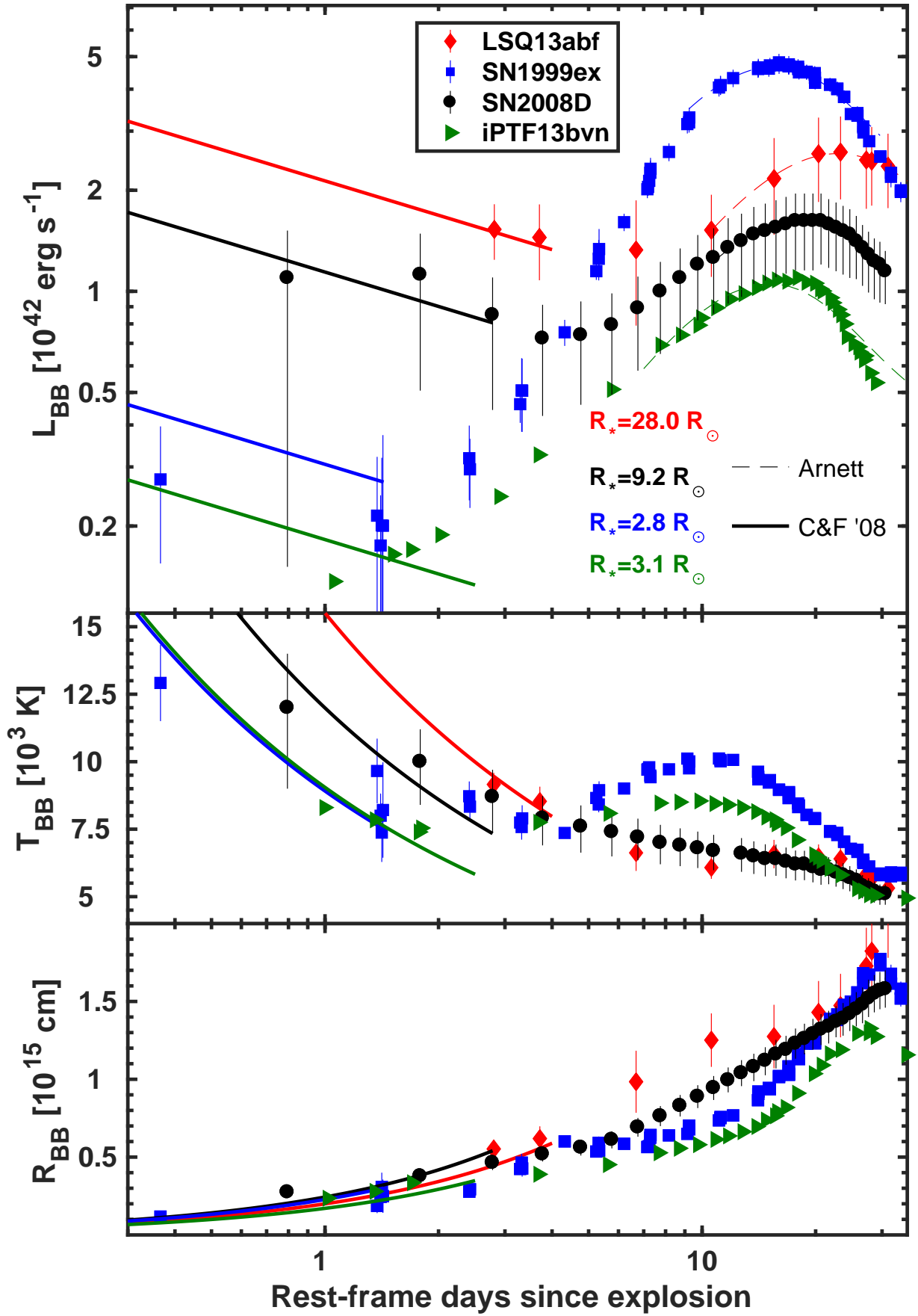}
\caption{\label{sbo_comp} The bolometric properties of SNe 1999ex, 2008D, iPTF13bvn, and LSQ13abf reproduced by the combined  Arnett  and \citeauthor{chevalier08} model discussed in Sect.~\ref{sec:model}. The derived progenitor radii are reported in the top panel.} 
\end{figure*}

How do these explosion parameters compare to those of  the progenitors of our small comparison sample of  SNe~Ib with documentation of an early peak? 
To answer this question we apply the  combined \citet{chevalier08} and \citeauthor{arnett82} model to the bolometric properties of   SN~1999ex, SN~2008D and iPTF13bvn as obtained from \citet{stritzinger02}, \citet{modjaz09}  and \citet{fremling16}, respectively. The model fits to the early peak and the photospheric phase  $L_{BB}$, $T_{BB}$, and $R_{BB}$ profiles of SN~1999ex, SN~2008D, iPTF13bvn, and LSQ13abf are shown in Fig.~\ref{sbo_comp} and the best-fit model parameters are summarized in Table~\ref{tab:litcomp}.
To compute these fits it was assumed that the BB velocity tracks the photospheric velocity, since the  BB velocity and  $v_{Fe}$ were found to be very similar for iPTF13bvn \citep{fremling14}.  
Inspection of Fig.~\ref{sbo_comp} reveals that  the post-SBO  cooling model reproduces the bolometric properties during the post-SBO cooling phase and the photospheric phase for each of the comparison objects. 

With values of  $R^{99ex}_{*} = 2.8\pm1.3(0.2)$~$R_{\odot}$, $R^{08D}_{*} = 9.2\pm1.9(0.8)$~$R_{\odot}$ and $R^{13bvn}_{*} = 3.1\pm1.9(0.2)$~$R_{\odot}$,  the \citet{chevalier08}  model suggests their progenitors are more compact as compared to that of LSQ13abf with an inferred value of $R^{13abf}_{*} = 28.0\pm3.3(6.7)$~$R_{\odot}$. 
Interestingly, with a combined model fit providing $M^{13abj}_{ej} \sim 6.0$ $M_{\odot}$, LSQ13abf also appears to originate from a more massive progenitor relative to not only our comparison sample with early peak, but also  relative to  average values ($\lesssim 4 M_{\odot}$) inferred from various SN~Ib sample studies (see Fig.~25 in \citealp{taddia18csp}).
These findings are consistent with LSQ13abf being both significantly brighter than the comparison sample during the post-SBO phase and exhibiting a broader light curve during its rise and decline to and from maximum.

\subsection{LSQ13abf progenitor scenario}
\label{sec:progenitor}

As previously mentioned, the majority of SE SNe likely originate from relatively low-mass progenitors, well below the canonical mass (i.e, ZAMS mass $\sim 25 M_{\odot}$) where it is possible for solitary massive stars to shed their hydrogen envelopes via steady-state line-driven winds over their evolutionary lifetimes.\footnote{Depending on adopted mass-loss models as well as stellar metallicity and rotation, the minimum ZAMS mass required for single stars to drive robust enough line-driven winds to strip the hydrogen envelope  could be as much as $40 M_{\sun}$ \citep[see, e.g.,][for a discussion and references therein]{smith11}.}
Therefore, many of the progenitors of SE SNe are thought to be in interacting  binaries that shed their hydrogen envelopes through Roche-lobe overflow  to a companion.
Both the Arnett and hydrodynamical models of LSQ13abf suggest a  large ejecta mass for a SN~Ib (5.9$-$6.2 $M_{\odot}$), which is compatible with a ZAMS mass of  $\sim 25 M_{\odot}$. 
This suggests that LSQ13abf is among the more massive SN~Ib yet studied.
Contrary to most other SE SNe that have less-massive progenitors and are likely in an interacting binary systems, the  high-mass inferred for LSQ13abf may imply  it is associated with a single star.
However, we note that massive stars can also be members of binary star systems \citep[see, e.g.,][]{yoon15}.
In short, it is difficult for us to ascertain if its origins lie in a  single or binary progenitor.

Comparison of the early observations of LSQ13abf to the hydrodynamical models of \citet{bersten13} and \citet{dessart18}  also suggest the presence of an extended envelope (i.e., $R_{*}$ $\sim$ 100~$R_{\odot}$). 
However, these models are not entirely consistent with LSQ13abf as we not only favor a He-star progenitor with an extended envelope, but as mentioned above, a progenitor more massive than  considered in these models and others in the literature.

Hydrogen-rich single massive stars close to the Eddington limit can also produce inflated envelopes \citep{ishii99, petrovic06, Sanyal15, fuller17}, although their $R_{*}$ are generally predicted to be smaller than those of relatively low mass He-giant stars in binary systems. In principle, similar to the H-rich scenario, solitary massive He stars near the Eddington limit could also suffer significant envelope inflation \citep{grafener12}.  To our knowledge there  are no observations or models in the literature of massive ($> 25 M_{\odot}$) He stars with significantly extended envelopes.  However, \citet{fuller18}  presented a simulation of a $5~M_{\odot}$ He-star  evolved from a $15~M_{\odot}$ ZAMS star, stripped of hydrogen through companion interaction, and during its pre-SN evolution significantly inflates its envelope via a super-Eddington wind driven by energy thermalized in the outer layers  that ultimately originates from deep within the core of the star and transferred by  internal gravity waves \citep{quataert12}. 
Along these lines, we  suggest that the progenitor of LSQ13abf might have been a  single or a binary, massive He-star with a significantly inflated envelope. Although beyond the scope of this paper, we do encourage others in the community to further investigate such progenitors.

In the case of SN~2008D, \citet{soderberg08} favored a massive WR progenitor star, while
\citet{modjaz09} concluded $R^{08D}_{*} \approx 12~R_{\odot}$ would be consistent with the typical 
radius of a massive WN star. However,  \citet{dessart18} argued that the progenitor was a low-mass He-giant star in a binary system.
While a low-mass He-giant star in a binary system would naturally produce the significant emission observed at  early epochs of SN~2008D and LSQ13abf, it is not consistent with the high ejecta mass we obtained for both objects (see Table~\ref{tab:litcomp}).

Regardless of the binary or single nature of the progenitor system, we find  the radius of the precursor was larger than that of the other SNe~Ib with published observations documenting an early peak. In a post-SBO cooling scenario with a regular progenitor structure, as modelled using \citet{chevalier08}, LSQ13abf's progenitor star at the moment of explosion had  $R^{13abf}_{*} \sim 28~R_{\odot}$. This is three times $R^{08D}_{*}$ assuming the same model. If we consider the extended-envelope structure, as suggested by the hydrodynamical models in \citet{bersten13} and \cite{dessart18}, then  $R^{13abf}_{*}$ might be even   larger compared to that of SN~2008D, perhaps on the order of $R^{13abf}_{*} \sim 100~R_{\odot}$.

\section{Conclusion}
\label{sec:conclusions} 

We presented early photometry and spectroscopy of the SN~Ib LSQ13abf, which shows an early peak in the bluer bands as in a few other cases in the literature.  Based on both semi-analytical and hydrodynamical models from the literature, we find  that the progenitor star of LSQ13abf had a larger radius than those of SNe~Ib including SN~1999ex, SN~2008D and iPTF13bvn,  all of which were observed during the post-SBO cooling phase. Hydrodynamical models of He stars in interacting binaries suggests a progenitor structure with an extended envelope attached to a dense core. 
The  high $M_{ej}$ estimate
points toward LSQ13abf originating from a  massive pre-SN progenitor  that likely evolved from a  $M_{ZAMS}$  $\gtrsim 25 M_{\odot}$  star. 
Such a progenitor could be in a binary or single star system.
No matter the exact nature of the progenitor, modern simulations of high mass stars do not exhibit extended envelopes as inferred for LSQ13abf.
We therefore hope that this study provides impetus for others to further explore   scenarios of single, massive He-stars with inflated envelopes that might ultimately produce objects similar to LSQ13abf.

\begin{acknowledgements} 
A special thank you to the referee for their constructive report.
We thank Takashi Jose Moriya, Thomas Tauris, and Paolo Mazzali for fruitful discussions. We acknowledge Jeffrey Silverman for reducing the HET spectroscopic observations, and Peter Nugent and David Rabinowitz for providing LSQ images. M.S., F.T., and E.K. are funded by a project grant (8021-00170B) from the Independent Research Fund Denmark.
 M.S. and S.H. are supported in part by a generous grant (13261) from VILLUM FONDEN. 
E.B. acknowledges support from NASA Grant: NNX16AB25G.
 N.B.S. acknowledges support from the NSF through grant AST-1613455, and through the Texas A\&M University Mitchell/Heep/Munnerlyn Chair in Observational Astronomy.
L.G. is supported by the European Union's Horizon 2020 research and innovation programme under the Marie Sk\l{}odowska-Curie grant agreement No. 839090.
 The CSP-II has been funded by the USA's NSF under grants AST-0306969, AST-0607438, AST-1008343, AST-1613426, AST-1613455, and AST-1613472, and in part by  a Sapere Aude Level 2 grant funded by the Danish Agency for Science and Technology and Innovation  (PI M.S.).
The research of JCW is supported in part by NSF AST-1813825. This work is partly based on observations made with the Nordic Optical Telescope, operated by the Nordic Optical Telescope Scientific Association at the Observatorio del Roque de los Muchachos, La Palma, Spain, of the Instituto de Astrofisica de Canarias. 
Visual-wavelength spectra of LSQ13abf were obtained in part with ALFOSC, which is provided by the Instituto de Astrofisica de Andalucia (IAA) under a joint agreement with the University of Copenhagen and NOTSA.
\end{acknowledgements}

\bibliographystyle{aa}

\clearpage

\onecolumn

\begin{deluxetable}{ccccccc}
\tablecolumns{9}
\tablewidth{0pt}
\tablecaption{Optical photometry of the local sequences for LSQ13abf in the {\em standard} system.\tablenotemark{a}\label{tab:lsq13abf_opt_locseq}}
\tablehead{
\colhead{ID} &
\colhead{$\alpha (2000)$} &
\colhead{$\delta (2000)$} &
\colhead{$r'$} &
\colhead{$i'$} &
\colhead{$B$}  &
\colhead{$V$}}
\startdata
2     & 177.23423  & 19.13506   &  $\cdots$          &   $\cdots$         &    14.077(0.012)  &   $\cdots$          \\    
3     & 177.24424  & 19.11977   &   14.272(0.017)    &    14.113(0.027)   &    14.957(0.016)  &    14.438(0.014)  \\  
4     & 177.27493  & 19.22754   &   15.441(0.016)    &    15.296(0.017)   &    16.080(0.029)  &    15.601(0.016)  \\  
5     & 177.31714  & 19.16629   &   16.183(0.026)    &    16.060(0.025)   &    16.851(0.035)  &    16.346(0.027)  \\  
6     & 177.29127  & 19.10672   &   17.884(0.054)    &    17.603(0.083)   &    19.110(0.056)  &    18.279(0.095)  \\
7     & 177.25087  & 19.20356   &   18.177(0.059)    &    17.996(0.098)   &    19.107(0.157)  &    18.458(0.070)  \\    
9     & 177.28277  & 19.19292   &   18.761(0.100)    &    18.704(0.103)   &    19.313(0.143)  &    18.861(0.130)  \\  
13    & 177.31741  & 19.11532   &   19.027(0.087)    &    18.759(0.156)   &    19.640(0.158)  &    19.184(0.127)  \\      
14    & 177.34702  & 19.11574   &   19.029(0.141)    &    18.557(0.168)   &    $\cdots$       &    19.618(0.158)  \\ 
\enddata
\tablenotetext{a}{Values in parenthesis are 1-$\sigma$ uncertainties and correspond to an rms of the instrumental errors of the photometry obtained over a minimum of three nights.}
\end{deluxetable}

\begin{deluxetable}{ccccccccc}
\tablecolumns{9}
\tablewidth{0pt}
\tablecaption{NIR photometry of the local sequences for LSQ13abf in the {\em standard} system.\tablenotemark{a}\label{tab:lsq13abf_nir_locseq}}
\tablehead{
\colhead{ID} &
\colhead{$\alpha (2000)$} &
\colhead{$\delta (2000)$} &
\colhead{$Y$} &
\colhead{N}   & 
\colhead{$J$} &
\colhead{N}   & 
\colhead{$H$} &
\colhead{N}   }

\startdata
101 & 177.25798 & 19.14391 & 17.270(047) &	1 & 16.899(042) & 3        & 16.434(111) &	3         \\         
102 & 177.30806 & 19.14254 & 17.342(062) &	1 & 17.666(042) & 1        & 16.530(198) &	2         \\
103 & 177.28269 & 19.19284 & 17.937(123) &	4 & 18.039(093) & 3        & 17.262(093) &	1         \\
104 & 177.29263 & 19.18621 & 18.482(349) &	2 & 18.231(012) & 2        & 17.338(072) &	1         \\
105 & 177.25755 & 19.16376 & 18.893(137) &	1 & 18.175(152) & 2        & 17.263(147) &	1         \\
107 & 177.27550 & 19.14922 & 17.517(069) &	1 & $\cdots$    & $\cdots$ & $\cdots$    &	$\cdots$   \\
\enddata
\tablenotetext{a}{Values in parenthesis are 1-$\sigma$ uncertainties computed by taking the weighted average of the instrumental errors of the photometry obtained during the nights photometric standard fields were observed. N represents the number of photometric nights that the local sequence stars were calibrated relative to standard field observations.}
\end{deluxetable}

\begin{deluxetable}{cccccccc}
\tabletypesize{\scriptsize}
\tablewidth{0pt}
\tablecaption{CSP-II Optical photometry of LSQ13abf in the \textit{natural} system.\label{tab:optphot}}
\tablehead{
\colhead{JD}&
\colhead{$B$}&
\colhead{JD}&
\colhead{$V$}&
\colhead{JD}&
\colhead{$r$}&
\colhead{JD}&
\colhead{$i$}\\
\colhead{(days)}&
\colhead{(mag)}&
\colhead{(days)}&
\colhead{(mag)}&
\colhead{(days)}&
\colhead{(mag)}&
\colhead{(days)}&
\colhead{(mag)}}
\startdata
      2456398.67  & 18.795(0.016) &   2456398.66 & 18.502(0.016)  &  2456398.66 &  18.478(0.017) &  2456398.65 & 18.619(0.021) \\
      2456399.58  & 18.855(0.015) &   2456399.59 & 18.482(0.015)  &  2456399.60 &  18.474(0.015) &  2456399.60 & 18.563(0.017) \\
      2456402.62  & 19.143(0.027) &   2456402.62 & 18.469(0.021)  &  2456402.61 &  18.329(0.019) &  2456402.61 & 18.425(0.021) \\
      2456406.60  & 19.107(0.079) &   2456406.60 & 18.418(0.100)  &  2456406.61 &  18.131(0.022) &  2456406.62 & 18.156(0.061) \\
      2456411.64  & 18.610(0.020) &   2456411.63 & 17.959(0.014)  &  2456411.62 &  17.799(0.013) &  2456411.61 & 17.850(0.013) \\
      2456416.58  & 18.442(0.012) &   2456416.58 & 17.786(0.013)  &  2456416.57 &  17.613(0.013) &  2456416.57 & 17.633(0.015) \\
      2456419.57  & 18.443(0.013) &   \ldots     & \ldots.        &  2456419.58 &  17.573(0.012) &  2456419.59 & 17.620(0.013) \\
      2456423.62  & 18.672(0.014) &   2456423.62 & 17.914(0.012)  &  2456423.63 &  17.614(0.012) &  2456423.64 & 17.607(0.013) \\
      2456424.55  & 18.744(0.015) &   2456424.56 & 17.945(0.013)  &  2456424.57 &  17.637(0.013) &  2456424.57 & 17.586(0.013) \\
      2456427.56  & 18.944(0.024) &   2456427.57 & 18.052(0.018)  &  2456427.55 &  17.704(0.014) &  2456427.55 & 17.586(0.017) \\
      2456434.63  & 19.414(0.064) &   2456434.62 & 18.355(0.026)  &  2456434.61 &  17.961(0.016) &  2456434.61 & 17.758(0.018) \\
      2456435.54  & 19.511(0.049) &   2456435.53 & 18.433(0.021)  &  2456435.52 &  18.002(0.017) &  2456435.52 & 17.773(0.016) \\
      2456436.62  & 19.558(0.066) &   2456436.61 & 18.496(0.024)  &  2456436.59 &  18.028(0.019) &  2456436.60 & 17.835(0.019) \\
      2456437.55  & 19.591(0.064) &   2456437.54 & 18.497(0.023)  &  2456437.53 &  18.038(0.019) &  2456437.53 & 17.821(0.019) \\
      2456438.55  & 19.561(0.059) &   2456438.54 & 18.570(0.026)  &  2456438.52 &  18.085(0.016) &  2456438.53 & 17.905(0.019) \\
      2456444.50  & 19.970(0.028) &   2456444.51 & 18.787(0.015)  &  2456444.52 &  18.345(0.014) &  2456444.53 & 18.122(0.016) \\
      2456445.51  & 19.975(0.025) &   2456445.50 & 18.823(0.015)  &  2456445.49 &  18.352(0.014) &  2456445.49 & 18.104(0.014) \\
      2456446.52  & 19.962(0.025) &   2456446.51 & 18.837(0.016)  &  2456446.52 &  18.385(0.016) &  2456446.53 & 18.134(0.015) \\
      2456447.53  & 20.029(0.024) &   2456447.51 & 18.866(0.015)  &  2456447.49 &  18.392(0.018) &  2456447.50 & 18.149(0.021) \\
      2456448.50  & 20.037(0.024) &   2456448.48 & 18.885(0.015)  &  2456448.46 &  18.433(0.013) &  2456448.47 & 18.182(0.015) \\
\enddata
\end{deluxetable}

\begin{deluxetable}{cccc}
\tabletypesize{\scriptsize}
\tablewidth{0pt}
\tablecaption{NIR photometry of LSQ13abf.\label{tab:nirphot}}
\tablehead{
\colhead{JD}&
\colhead{$Y$}&
\colhead{$J$}&
\colhead{$H$}\\
\colhead{(days)}&
\colhead{(mag)}&
\colhead{(mag)}&
\colhead{(mag)}}
\startdata
2456400.64  & 17.587(041) &  17.455(028) & 17.407(064)  \\ 
\enddata
\end{deluxetable}

\begin{deluxetable}{ccccccccc}
\tablecolumns{9}
\tablewidth{0pt}
\tablecaption{Optical photometry of the local sequence used in relation with LSQ imaging of LSQ13abf\label{LSQlocseq}}
\tablehead{
\colhead{ID} &
\colhead{$\alpha (2000)$} &
\colhead{$\delta (2000)$} &
\colhead{$V_{LSQ}$} &
\colhead{$V_{nat}$} &
\colhead{$r_{std}$} &
\colhead{$i_{std}$} &
\colhead{$B_{std}$} &
\colhead{$V_{std}$}}
\startdata
201 & 11:49:07.96 & +19:11:34.8   & 18.927(096) &    18.861(095) & 18.762(071) & 18.711(080) & 19.342(103) & 18.853(095) \\
202 & 11:49:00.31 & +19:12:12.7   & 18.455(051) &    18.458(051) & 18.185(045) & 18.001(072) & 19.151(111) & 18.433(051) \\
204 & 11:49:16.23 & +19:09:58.0   & 16.369(017) &    16.347(017) & 16.181(017) & 16.057(015) & 16.885(018) & 16.331(017) \\
205 & 11:49:02.14 & +19:08:37.9   & 19.737(098) &    19.778(096) & 19.245(082) & 18.409(091) &   $\cdots$     & 19.703(096) \\
207 & 11:49:21.71 & +19:14:13.0   & 19.071(084) &    18.991(083) & 18.555(053) & 18.081(066) &   $\cdots$     & 18.941(083) \\
210 & 11:49:28.91 & +19:10:16.6   & 17.375(049) &    17.424(049) & 17.120(034) & 16.987(031) &   $\cdots$     & 17.400(049) \\
211 & 11:49:24.90 & +19:11:51.0   & 19.383(087) &    19.351(086) & 19.015(075) & 18.484(063) &    $\cdots$     & 19.303(086) \\
218 & 11:48:54.18 & +19:10:33.7   & 19.643(055) &    19.692(052) & 19.027(131) & 17.657(057) &    $\cdots$     & 19.580(052) \\
221 & 11:48:58.38 & +19:14:35.5   & 18.626(059) &    18.660(058) & 18.427(080) & 18.231(051) & 19.397(092) & 18.636(058) \\
230 & 11:49:25.75 & +19:09:36.3   & 16.391(023) &    16.417(023) & 16.091(006) & 15.866(025) & 17.164(052) & 16.387(023) \\
\enddata
\end{deluxetable}

\begin{deluxetable}{cc}
\tablewidth{0pt}
\tablecaption{LSQ $V$-band photometry  of LSQ13abf.\label{tab:lsq}}
\tablehead{
\colhead{JD}&
\colhead{$V_{LSQ}$}\\
\colhead{(days)}&
\colhead{(mag)}}
\startdata
2456397.526 & 18.622(0.016) \\  
2456397.645 & 18.572(0.016) \\  
2456397.727 & 18.568(0.016) \\  
2456403.509 & 18.560(0.059) \\  
2456403.593 & 18.500(0.030) \\  
2456407.517 & 18.255(0.037) \\  
2456407.601 & 18.305(0.057) \\  
2456409.596 & 18.336(0.150) \\  
2456411.512 & 17.959(0.018) \\  
2456411.595 & 18.021(0.034) \\  
\enddata
\end{deluxetable}

\begin{deluxetable}{cccccc}
\tablewidth{0pt}
\tablecaption{Spectroscopy of LSQ13abf\label{tab:spectra}}
\tablehead{
\colhead{Date (UT)}&
\colhead{JD-2,456,000}&
\colhead{Phase\tablenotemark{a}}&
\colhead{Telescope}&
\colhead{Instrument}&
\colhead{Range}\\
\colhead{}&
\colhead{(days)}&
\colhead{(days)}&
\colhead{}&
\colhead{}&
\colhead{(\AA)}}
\startdata
16 Apr 2013    &   399.36    &  +3.5   & NOT   &   ALFOSC   & 3400--9150     \\
14 May 2013    &   427.46    &  +31.0   & NOT   &   ALFOSC   & 3500--9140     \\
20 May 2013    &   432.58    &  +36.0   & Baade   &   FIRE   & 8300--25000	     \\
20 May 2013    &   432.72    &  +36.2   & HET   &   LRS   & 4172--10800     \\
27 May 2013    &   440.46    &  +43.8   & NOT   &   ALFOSC   & 3400--9140     \\
13 Jun 2013    &    457.46    &  +60.4   & NOT   &   ALFOSC   & 3350--9140     \\
\enddata
\tablenotetext{a}{From explosion date, in rest-frame.}
\end{deluxetable}

\begin{deluxetable}{lcc}
\tablecolumns{3}
\tablewidth{0pt}
\tablecaption{Host-galaxy line fluxes as measured by the gaussian fits shown in Fig.~\ref{hostspec}.\label{tab:host_line_fluxes}}
\tablehead{
\colhead{line} &
\colhead{Flux} &
\colhead{Flux uncertainty} \\
\colhead{} &
\colhead{(10$^{-16}$ erg~s$^{-1}$~cm$^{-2}$)} &
\colhead{(10$^{-16}$ erg~s$^{-1}$~cm$^{-2}$)}}
\startdata
H$\beta~\lambda$4861         & 1.64 & 0.04 \\
$[$\ion{O}{iii}$]~\lambda$4959  & 6.72 & 0.02 \\
$[$\ion{O}{iii}$]~\lambda$5007  & 1.85 & 0.02 \\
H$\alpha~\lambda$6563        & 5.93 & 0.09 \\
$[$\ion{N}{ii}$]~\lambda$6584   & 1.51 & 0.03 \\
$[$\ion{S}{ii}$]~\lambda$6717   & 1.39 & 0.03 \\
$[$\ion{S}{ii}$]~\lambda$6731   & 1.05 & 0.02 \\
\enddata
\end{deluxetable}

\begin{deluxetable}{ccc}
\tabletypesize{\scriptsize}
\tablewidth{0pt}
\tablecaption{Peak epochs and magnitudes of the light curves of LSQ13abf.\label{tab:peak}}
\tablehead{
\colhead{Band}&
\colhead{Peak epoch}&
\colhead{Peak magnitude} \\
\colhead{}&
\colhead{(MJD)}&
\colhead{(mag)}}
\startdata
$B$ & 56416.88(0.24) & 18.426(0.009)\\
$V$ & 56417.30(0.33) & 17.789(0.012)\\
$r$ & 56419.41(0.38) & 17.576(0.008)\\
$i$ & 56423.76(1.57) & 17.591(0.007)\\
\enddata
\end{deluxetable}

\begin{deluxetable}{ccccc}
\tabletypesize{\scriptsize}
\tablewidth{0pt}
\tablecaption{Velocity of the spectral lines of LSQ13abf\label{tab:vel}}
\tablehead{
\colhead{Date (UT)}&
\colhead{JD-2,456,000}&
\colhead{Phase\tablenotemark{a}}&
\colhead{Vel ($\ion{Fe}{ii}$ $\lambda$5169)} &
\colhead{Vel ($\ion{He}{i}$ $\lambda$5876)}\\
\colhead{}&
\colhead{(days)}&
\colhead{(days)}&
\colhead{(km~s$^{-1}$)}&
\colhead{(km~s$^{-1}$)}}
\startdata
14 May 2013    &   427.46    &  +31.0   &  $6400\pm100$     & $9100\pm600$  \\
20 May 2013    &   432.72    &  +36.2   &  $5300\pm1000$ & $900\pm800$ \\   
27 May 2013    &   440.46    &  +43.8   &   $5100\pm300$ & $8600\pm200$\\
13 Jun 2013    &   457.46    &  +60.4    &  $5300\pm300$ & $7000\pm300$ \\
\enddata
\tablenotetext{a}{Rest-frame days post estimated explosion epoch.}
\end{deluxetable}

\begin{deluxetable}{c|cccc}
\tabletypesize{\scriptsize}
\tablewidth{0pt}
\tablecaption{Explosion and progenitor parameters for SN~1999ex, SN~2008D, iPTF13bvn, and LSQ13abf from the semi-analytic modelling shown in Fig.~\ref{sbo_comp}.\label{tab:litcomp}}
\tablehead{
\colhead{SN}&
\colhead{$M_{ej}$}&
\colhead{$E_K$}&
\colhead{$^{56}$Ni mass}&
\colhead{$R_*$}\\
\colhead{}&
\colhead{($M_{\odot}$)}&
\colhead{(10$^{51}$ erg)}&
\colhead{($M_{\odot}$)}&
\colhead{($R_{\odot}$)}}
\startdata
SN~1999ex   &   3.04$\pm$0.08(0.63)    &  1.09$\pm$0.03(0.10)   & 0.21$\pm$0.01(0.03)  & 2.8$\pm$1.3(0.2)  \\
SN~2008D    &   4.56$\pm$0.14(0.95)    &  1.74$\pm$0.06(0.36)   & 0.08$\pm$0.01(0.03)  & 9.2$\pm$1.9(0.8)  \\
iPTF13bvn    &   2.08$\pm$0.17(0.43)    &  0.40$\pm$0.03(0.08)  & 0.05$\pm$0.01(0.02)   & 3.1$\pm$1.9(0.2)  \\
LSQ13abf    &   5.94$\pm$0.14(1.09)     &  1.27$\pm$0.04(0.23)   & 0.16$\pm$0.01(0.02)  & 28.0$\pm$3.3(6.7)  \\
\enddata
\tablecomments{The errors outside the parentheses correspond to  the fit error, while the errors between parentheses are obtained assuming an uncertainty of  18\%\ of the photospheric velocity, which affects the ratio between kinetic energy and ejecta mass, and a   7\% uncertainty on the distance. The uncertainty in parentheses for the $^{56}$Ni mass estimates also  include an  additional   10\% uncertainty added in quadrature to be conservative. }
\end{deluxetable}

\end{document}